\documentclass{iopjournal}
\usepackage{graphicx} 

\usepackage[T1]{fontenc}
\usepackage{lmodern}

\usepackage{amsmath}
\usepackage{amsfonts}
\usepackage{amssymb}
\usepackage{pifont}
\usepackage{ragged2e} 

\newcommand{\cwinpy}{\texttt{CWInPy}}
\newcommand{\weave}{\texttt{Weave}}
\newcommand{\pyfstat}{\texttt{PyFstat}}
\newcommand{\pfa}{p_{\rm fa}}
\newcommand{\Fstat}{\mathcal{F}}
\newcommand{\Tcoh}{T_{\rm coh}}
\newcommand{\Nseg}{N_{\rm seg}}

\newcommand{\cmark}{\ding{51}}%
\newcommand{\xmark}{\ding{55}}%

\newcommand{\UIB}{IAC3, Universitat de les Illes Balears, Cra.~de Valldemossa km 7.5, 07122 Palma, Spain}
\newcommand{\INFNRM}{INFN, Sezione di Roma, Piazzale Aldo Moro, 2, I-00185 Roma, Italy}
\newcommand{\INFNFE}{INFN, Sezione di Ferrara, Via Saragat 1, 44122 Ferrara, Italy}
\newcommand{\CAMK}{Nicolaus Copernicus Astronomical Center, Polish Academy of Sciences, Bartycka 18, 00-716 Warsaw, Poland}
\newcommand{\Sapienza}{Universit\`a di Roma La Sapienza, Piazzale Aldo Moro, 2, I-00185 Roma, Italy}
\newcommand{\UMich}{Department of Physics, University of Michigan, 450 Church St, Ann Arbor, MI 48109}

\newcommand{\RESCEU}{Research Center for the Early Universe (RESCEU), Graduate School of Science, The University of Tokyo, Tokyo 113-0033, Japan}

\newcommand{\UniMinn}{University of Minnesota: Minneapolis, Minnesota, US}

\newcommand{\ozgrav}{Australian Research Council Centre of Excellence for Gravitational Wave Discovery (OzGrav), Hawthorn VIC 3122, Australia}

\newcommand{\MathPol}{Institute of Mathematics,  Polish Academy of Sciences, Warsaw, Poland and National Centre for Nuclear Research, Swierk, Poland}

\newcommand{\SchoolPhys}{School of Physics, University of Melbourne, Parkville VIC 3010 Australia}

\newcommand{\SouthamptonUni}{University of Southampton, Southampton SO17 1BJ, United Kingdom}

\begin{document}

\justifying 

\articletype{Paper} 

\title{The G347.3\textminus0.5 outlier from O3: a follow-up case study for continuous gravitational-wave candidates}

\fancyhead[L]{\small Authors' preprint (2026): G347.3\textminus0.5 follow-up case study}
\fancyhead[R]{\small L. Mirasola \textit{et al.}}

\author{L. Mirasola\orcid{0000-0003-1606-4183}$^1$,
F. Amicucci\orcid{0009-0005-2139-4197}$^{2,\,3}$,
M. Carrio\orcid{0009-0009-7114-5067}$^1$,
D. H. T. Cheung\orcid{0000-0003-3905-0665}$^4$,
M. A. Ferrer-Martinez\orcid{0009-0008-9801-9506}$^{1,\,5}$,
E. Goetz\orcid{0000-0003-2666-721X}$^6$,
D. Keitel\orcid{0000-0002-2824-626X}$^1$,
A. M. Knee\orcid{0000-0003-0703-947X}$^4$,
P. Leaci\orcid{0000-0002-3997-5046}$^{2,\,3}$,
C. Palomba\orcid{0000-0002-4450-9883}$^3$,
K. Pham\orcid{0000-0002-7650-1034}$^7$,
O. J. Piccinni\orcid{0000-0001-5478-3950}$^1$,
M. Pitkin\orcid{0000-0003-4548-526X}$^{8,\,9}$,
I. Prohens\orcid{0009-0007-3075-075X}$^1$,
K. Riles\orcid{0000-0002-6418-5812}$^4$,
S. Safi-Harb\orcid{0000-0001-6189-7665}$^{10}$,
G. Woan\orcid{0000-0003-0381-0394}$^9$,
Z. Zhang,
M. Bejger\orcid{0000-0002-4991-8213}$^{11,\,12}$,
A. Calafat\orcid{0009-0008-7515-6305}$^{1}$,
R. Jaume\orcid{0000-0001-8691-3166}$^{1}$,
D. I. Jones$^{13}$,
A. Krolak\orcid{0000-0003-4514-7690}$^{14}$,
I. La Rosa\orcid{0000-0003-0107-1540}$^1$,
A. Melatos\orcid{0000-0003-4642-141X}$^{15,16}$,
J. R. Mérou\orcid{0000-0002-5776-6643}$^{1}$,
A. Nemmani\orcid{0009-0005-4620-7052}$^{10,\,11}$,
B. Rajbhandari\orcid{0000-0001-7568-1611}$^{17}$,
C. Salvadore\orcid{0009-0002-9967-4111}$^{2,3}$,
A. M. Sintes\orcid{0000-0001-9050-7515}$^{1}$,
R. Tenorio\orcid{0000-0002-3582-2587}$^{1}$,
K. Wette\orcid{0000-0002-4394-7179}$^{18,16}$,
J. T. Whelan\orcid{0000-0001-5710-6576}$^{19,20}$,
T. S. Yamamoto\orcid{0000-0002-8181-924X}$^{21}$
}

\affil{$^1$\UIB}

\affil{$^2$\Sapienza}

\affil{$^3$\INFNRM}

\affil{$^4$\UMich}

\affil{$^5$ Università di Milano and INFN sezione di Milano, Via Celoria 16, 20133 Milano, Italy}

\affil{$^6$University of British Columbia, Vancouver, BC V6T 1Z4, Canada}

\affil{$^7$\UniMinn}

\affil{$^8$CEDAR Audio Ltd, Cambridge CB21 5BS, United Kingdom}

\affil{$^9$IGR, University of Glasgow, Glasgow G12 8QQ, United Kingdom}

\affil{$^{16}$Department of Physics \& Astronomy, University of Manitoba, Winnipeg, MB R3T 2N2, Canada}

\affil{$^{11}$\INFNFE}

\affil{$^{12}$\CAMK}

\affil{$^{13}$\SouthamptonUni}

\affil{$^{14}$\MathPol}

\affil{$^{15}$\SchoolPhys}

\affil{$^{16}$\ozgrav}

\affil{$^{17}$Department of Physics, University of Maryland Baltimore County, 1000 Hilltop Circle, Baltimore, Maryland 21250, USA}

\affil{$^{18}$Centre for Gravitational Astrophysics, Australian National University, Canberra ACT 2601, Australia}

\affil{$^{19}$School of Physics and Astronomy and Center for Computational Relativity and Gravitation, Rochester Institute of Technology, Rochester, NY 14623, USA}

\affil{$^{20}$Department of Physics, University of Warwick, Gibbet Hill Road, Coventry CV4 7AL, United Kingdom}

\affil{$^{21}$\RESCEU}

\email{l.mirasola@uib.cat}

\begin{abstract}
We report a multi-pipeline follow-up study of the continuous gravitational-wave (CW) outlier identified by the Ming et al. Einstein{@}Home directed search for the young supernova remnant G347.3\textminus0.5 (containing the central compact object RXJ1713.7\textminus3946).
The outlier was initially identified in LIGO data from the O3a observing run and followed up with the addition of O3b and O4a data by the Einstein{@}Home pipeline, with O4a producing weaker evidence.
Here, we show extended data quality checks and results from three pipelines that have independently investigated this outlier, including the newly released O4b data set. 
We can robustly recover the outlier in O3, while we find no evidence for a related standard CW signal in O4a (consistent with Ming et al.) and O4b data.
This work serves as a useful test case for future multi-pipeline follow-ups of CW candidates that aim to investigate their astrophysical or noise nature, and it is therefore an important step in preparation for the first CW detection.
\end{abstract}

\section{Introduction}

Continuous gravitational waves (CWs) are persistent gravitational waves expected to originate from rapidly rotating neutron stars (NSs) with non-axisymmetric mass distributions~\cite{Riles:2022wwz}.
While no CWs have yet been detected, their observation would provide valuable insights into the population and properties of Galactic NSs~\cite{Jones:2024npg,Owen:2025ata},
which to date have been observed exclusively through electromagnetic observations~\cite{Manchester:2004bp,Reed:2021scb}.

Among many possible astrophysical sources, young core-collapse supernova remnants (SNRs) are of special interest for CW searches.
They can host young NSs, which might be able to sustain larger deformations than older stars~\cite{LIGOScientific:2008hqb}.
Thanks to multi-wavelength observations, parameters such as sky location and source distance are well constrained~\cite{Ferrand:2012jh}.
As a result, CW searches can focus on the region of interest and rely on longer integration times (usually referred to as coherence times, $\Tcoh$) compared to wider parameter space searches.
However, in the absence of pulsations, searches have to explore vast regions in the frequency and spin-down (time-derivatives of the rotational frequency) parameter space to identify a signal.
As a consequence, it is not feasible to analyse a full observing run via fully coherent approaches, as the required computational cost increases with a high power of the observing time~\cite{Wette:2023dom}.
Therefore, a hierarchical semi-coherent scheme is usually used, which trades sensitivity for a manageable computing cost.
After the initial search stage, interesting candidates are then followed up with more sensitive methods~\cite{Tenorio:2021wmz}.

A recent search of LIGO~\cite{LIGOScientific:2014pky} data by the Einstein@Home team (Ming et al.,~\cite{Ming:2025ehy}) has reported a low-significance CW outlier that survived several post-processing and follow-up stages, corresponding to the sky location of the central compact object (CCO) 1WGA J1713.4\textminus3949 in the SNR G347.3\textminus0.5 (also known as RX J1713.7\textminus3946).
On the other hand, recent results from the LIGO--Virgo--KAGRA (LVK) collaboration on O3a~\cite{LIGOScientific:2021mwx} and O4a data~\cite{LIGOScientific:2026hrn}\footnote{See~\cite{LIGOScientific:2014pky, VIRGO:2014yos, KAGRA:2020tym} for more information on the LVK detectors and \cite{KAGRA:2023pio,LIGOScientific:2025snk,LIGOScientific:2026jgl} for overviews of the data sets from the various observing runs.} on the same target resulted in upper limits that remain above the amplitude estimated for the outlier, with their search depth mostly limited by computing power constraints.
Consequently, those searches were not yet sensitive enough to either detect or confidently exclude a signal like the Einstein@Home outlier, providing the primary motivation for the present work.

Here, we use this outlier to detail the first coordinated effort to independently follow up a CW candidate that was reported from one search pipeline and survived initial follow-ups by the same team~\cite{Ming:2025ehy} with multiple and complementary analysis methods.
Beyond this specific case, such an effort will be crucial on the path towards the first CW detection, since establishing the astrophysical nature of an outlier requires independent confirmation across multiple methods.

The search in~\cite{Ming:2025ehy} focused on data from the third LVK observing run (O3)~\cite{KAGRA:2023pio} and also used data from the first part of the fourth observing run (O4a)~\cite{LIGOScientific:2025snk} for follow-up.
Here, we also include data from the recently released second part of O4 (O4b)~\cite{LIGOScientific:2026jgl}.
We report on data characterisation procedures used to check for possible correlations between the outlier and detector disturbances, extending the studies of~\cite{Ming:2025ehy}.
We also present results from a multi-pipeline follow-up using three distinct CW analysis packages: \cwinpy~\cite{cwinpy}, \weave~\cite{Wette2018}, and \pyfstat~\cite{Ashton:2018ure, Keitel:2021xeq, pyfstat}.
Finding consistent results among them ensures that our conclusions are less affected by single-pipeline specificities.
Ming et al. already reported reduced O4a evidence and unconvincing phase-parameter consistency.
We can confirm the outlier in O3 data, while neither O4a and O4b show evidence for a long-lived phase-coherent CW signal for parameters consistent with O3.

One possible explanation for a source appearing as a CW search outlier in O3 but not in the following run is that it does not follow the standard CW emission scenario. For instance, a NS experiencing a strong glitch~\cite{Antonopoulou:2022rpq,Ashton:2017wui} could undergo a sudden change in spin frequency and/or spin-down, causing the signal to fall outside the parameter space covered by the follow-up search or lose coherence with the assumed signal model.
We also tested a version of this hypothesis,
but an exhaustive study of possible deviations from the standard scenario is beyond the scope of this work.

The structure of the paper is as follows.
Section~\ref{sec:outlier} outlines the outlier and its main features, as reported by \cite{Ming:2025ehy}.
Section~\ref{sec:gw_data} describes the GW data used in this analysis
and summarises our data quality investigations for a narrow frequency range around the outlier.
Then, we present the checks for a CW signal near the candidate parameters using three separate methods in Sections~\ref{sec:o3_search} and \ref{sec:o4_search} using O3 and O4ab data, respectively.
Lastly, we draw our conclusions in Section~\ref{sec:conclusions}.

\section{The G347.3\textminus0.5 outlier\label{sec:outlier}}

In~\cite{Ming:2025ehy}, searches for CWs from three SNRs have been conducted on Einstein@Home.
The search for the G347.3\textminus0.5 SNR has produced a low-significance outlier with an estimated false-alarm probability of $\pfa\sim$10\%.
We report here its characteristics (Section~\ref{subsec:outlier_pars}) preceded by a summary of the considered signal model (Section~\ref{subsec:signal}).

\subsection{Signal model and detection framework\label{subsec:signal}}

The search in~\cite{Ming:2025ehy}, and this work, consider the standard CW model for signals emitted by an isolated NS.
The emitted GW frequency is proportional to the rotation frequency, with a model-dependent coefficient~\cite{Riles:2022wwz} (e.g., 2 if the star carries a mass quadrupole deformation and rotates around one of its principal axes), and evolves gradually as
\begin{equation}
    \label{eq:fcw_of_t}
    f(t) = f_0 + \dot f_0 (t-t_{\rm ref}) + \frac{1}{2} \ddot f_0 (t-t_{\rm ref})^2 + \frac{1}{6} \dddot f_0 \, (t-t_{\rm ref})^3 + ...
\end{equation}
with $f_0,\,\dot f_0,\, \ddot f_0, \dddot f_0$ the signal frequency and its time derivatives evaluated at the reference time $t_{\rm ref}$.
The search in~\cite{Ming:2025ehy} truncated the expansion at the second order.

The expected signal at the detector is~\cite{Jaranowski:1998qm}
\begin{equation}
    h(t)  = h_0\,F_+(t;\vec{n},\psi)\,  \frac{1+\cos^2\iota}{2}\, \cos \phi(t) +
        h_0\,F_\times(t;\vec{n},\psi)\, \cos\iota\,\sin \phi(t) ,
\end{equation}
where $h_0$ is the CW amplitude, $\iota$ the inclination angle of the source's rotation axis relative to the line of sight, and $\phi(t)$ the signal phase obtained by integrating Equation~\ref{eq:fcw_of_t}, after accounting for Doppler and relativistic effects due to the Earth's motion~\cite{Jaranowski:1998qm}.
The functions $F_{+,\times}$ are the time-dependent beam pattern functions encoding the detector response to a source located at sky position $\vec{n}$ with polarization angle $\psi$.

The search in~\cite{Ming:2025ehy} is based on the Einstein@Home distributed computing framework and hierarchical search pipeline~\cite{Brady:1997ji, Brady:1998nj, Pletsch:2008gc, Pletsch:2009uu, Pletsch:2010xb}.
After an initial stage using a semi-coherent line-robust detection statistic~\cite{Keitel:2013wga,Keitel:2015ova}, the detection statistic used in the follow-up stages, and hence for evaluating the outlier, is the maximum-likelihood multi-detector $\mathcal{F}$-statistic~\cite{Jaranowski:1998qm, Cutler:2005hc}.
This is obtained by matched filtering the data for a given combination of phase-evolution parameters $f_0,\,\dot f_0,\, \ddot f_0,\, \vec n$ (i.e., a template), analytically maximising over amplitude parameters.
The initial search and follow-up stages use a semi-coherent combination of $\mathcal{F}$-statistics from shorter segments, while the final stages use the fully-coherent $\mathcal{F}$-statistic over the full considered data set.

In this work, we generally quote values for the averaged semi-coherent $2\Fstat$-statistic.
Following, e.g., \cite{Wette:2021tbv},
we define
\begin{equation}
    2\hat\Fstat = \frac{1}{\Nseg} \sum_{i=1}^{\Nseg} 2\Fstat_i\,,
\end{equation}
where $2\Fstat_i$ is the statistic computed for segment $i$, and $\Nseg$ is the number of equal-length segments, each of duration $\Tcoh$, into which the data have been divided.
In stationary Gaussian noise, $2\hat\Fstat \times \Nseg$ is distributed as a $\chi^2$ with $4\Nseg$ degrees of freedom.
In the presence of a signal, it follows a non-central $\chi^2$ distribution with the same degrees of freedom and non-centrality parameter equal to the squared signal-to-noise ratio $\rho^2$.
For $\Nseg=1$, corresponding to the fully coherent case, we will simply refer to these as $2\Fstat$.

\subsection{The outlier and its parameters\label{subsec:outlier_pars}}
Recent estimates give an age for the G347.3$-$0.5 SNR of $\sim$ 1.0--1.7\,kyr~\cite{Mayer:2021wvs} and place it at a distance of $\sim$ 0.9--1.1\,kpc, see \texttt{SNRcat}~\cite{snrcat} and references therein.
While Ming et al.~\cite{Ming:2025ehy} also discussed the recent updates to these estimated ranges, in their analysis they adopted single values for the age and distance of 1.6\,kyr and 1.3\,kpc, respectively.
The value of 1.6\,kyr is based on the proposed association with the historical guest star of AD 393~\cite{1997A&A...318L..59W} -- an association that is plausible, but not firmly established.
Those quantities do not affect our analysis, only a possible astrophysical interpretation of the candidate
(as the age only enters the parameter-space setup of the first-stage wide search, and the distance only enters the signal model through the overall amplitude $h_0$ and is degenerate with NS ellipticity and moment of inertia).
The SNR's equatorial coordinates are $(\alpha_0,\delta_0) = \left(17^{\mathrm h}13^{\mathrm m}28.3^{\mathrm s},\, -39^\circ49'53.3''\right)$ with an uncertainty of $0.8''$ on both $\alpha$ and $\delta$ at the 99\% confidence level (CL)~\cite{Mignani:2008ew}.

The search for G347.3\textminus0.5 detailed in~\cite{Ming:2025ehy} divided the frequency band into low and high ranges, namely 20--500~Hz and 500--1500~Hz, optimizing its setup accordingly.
In total, $\mathcal{O}(10^{18})$ templates have been explored (see Figure~1 in~\cite{Ming:2025ehy}), with the outlier found in the low-frequency band.
The search initially considered O3a data to identify candidates, which were then subjected to a multi-stage follow-up procedure.
At the beginning, the G347.3\textminus0.5 low-frequency search focused on O3a data with $\Tcoh=60$ days, yielding $\sim 4.8 \times 10^{6}$ candidates to follow up.
Only those with $2\hat\Fstat$ consistent with simulations were further analysed in successive stages.
The low-frequency follow-up consisted of up to three steps, in which fully coherent searches with O3a-only, full-O3, and O4a-only data were considered, respectively. 

Out of the 13 candidates reaching the last follow-up stage, only one survives.
After accounting for the number of trials from the final stage of follow-ups, the authors report an associated $\pfa$ of about $10\%$.
The candidate's frequency and spin-down parameters from \cite{Ming:2025ehy}, reproduced here in Table~\ref{tab:G347cand}, together with the SNR sky location, are the starting point of this study.
Lastly, by performing an additional follow-up stage on O3 and O4a with the Bayesian method from~\cite{Ashok:2024fts}, the authors reported an estimated amplitude of the candidate $h_0\sim 6 \times 10^{-26}$, below previously reported upper limits~\cite{LIGOScientific:2026hrn, LIGOScientific:2021mwx, Salvadore:2025lba, McGloughlin:2025eso, Ming:2024dug, KAGRA:2022dwb, LIGOScientific:2026plm}.
However, the corresponding inclination angle was not reported, leaving open some ambiguity in the $h_0$ due to the strong degeneracy between the two parameters, as we will see in Section \ref{sec:o3cwinpy}.

\begin{table}[!ht]
    \centering
    \renewcommand*{\arraystretch}{1.4}
    \begin{tabular}{  c  c  c  c  c  c }
    \hline\hline
    $f_0$ [Hz] & $\dot{f}_0$ [Hz/s] & $\ddot{f}_0$ [Hz/s$^2$] &
    $\Delta^{99\%} f_0$ [Hz] & $\Delta^{99\%} \dot{f}_0$ [Hz/s] & $\Delta^{99\%} \ddot{f}_0$ [Hz/s$^2$] \\
    \hline
    $31.691150039$ & $-3.598467\times 10^{-10}$ & $7.744\times 10^{-20}$ &
    $\pm 1.5\times 10^{-8}$ & $\pm 5.0 \times 10^{-15}$ & $\pm 8.0 \times 10^{-22}$ \\
    \hline
    \end{tabular}
    \caption{
    \label{tab:G347cand}
    Parameters of the G347.3\textminus0.5 surviving candidate reported by~\cite{Ming:2025ehy} from their full-O3 follow-up stage.
    The parameters are reported at $t_{\rm ref} = 1246197626.5$~s (GPS).
    Uncertainties are quoted as 99\% credible intervals.
    }
\end{table}

The authors report a fully-coherent detection statistic in O3 of $2\Fstat_{\rm O3}\simeq 81$ in a joint-detector Bayesian follow-up analysis~\cite{Martins:2025jnq}, while their original final-stage fully-coherent O4a grid-based follow-up yielded $2\Fstat_{\rm O4a}=45$.
Bayesian results on O3 and O4a data yielded phase consistency that was deemed ``not convincing''.
Rerunning the O4a Bayesian analysis with a third-order frequency-derivative parameter included, they reported that the posteriors became more informative than their priors (taken as the O3 posteriors), but that the signal evidence still decreased compared to O3.
They also performed additional checks, including a cross-check with the O3 Einstein@Home and Falcon all-sky searches \cite{McGloughlin:2025eso, Dergachev:2025ead}, a preliminary O4a investigation with Falcon, and several data quality investigations (see next section).

\begin{figure}[!ht]
    \centering
    \includegraphics[width=\textwidth]{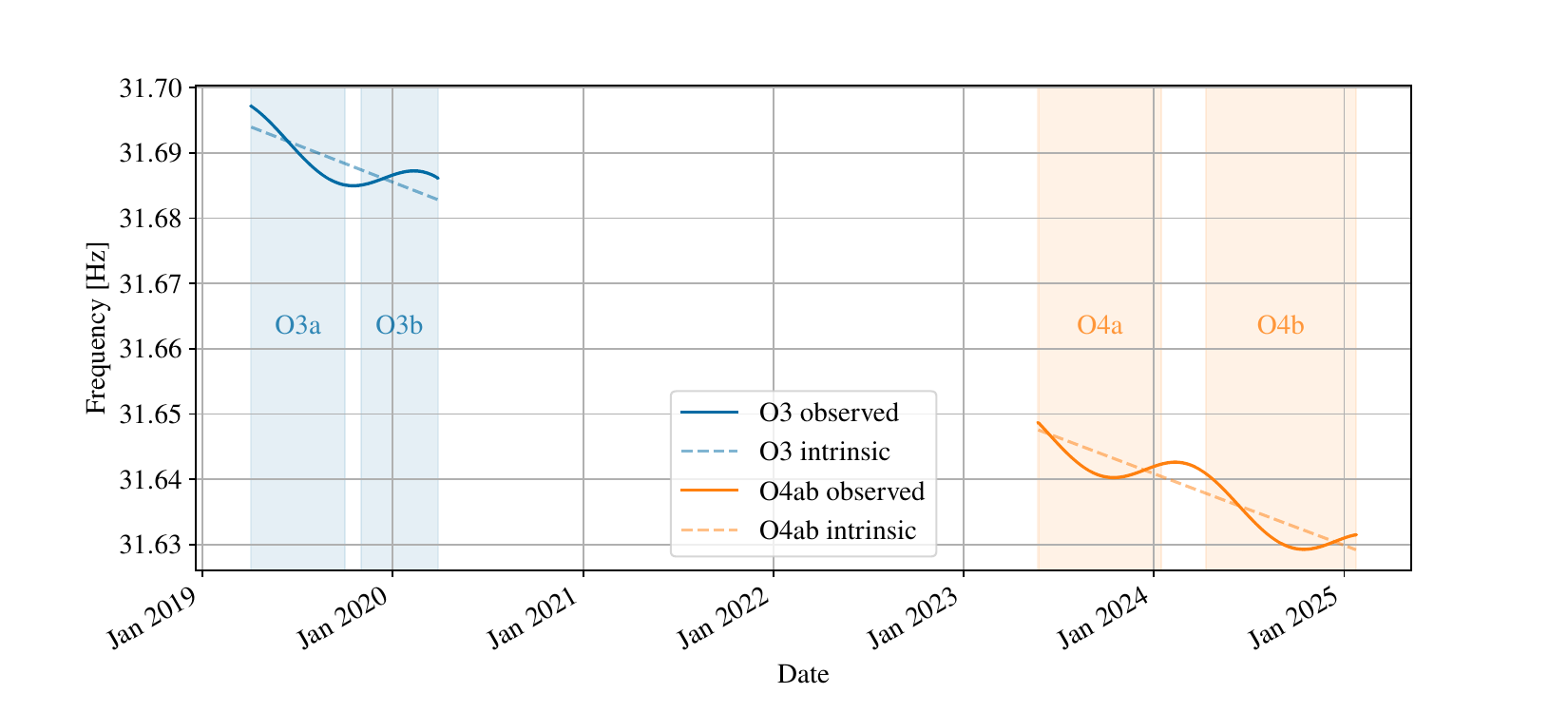}
    \caption{Frequency evolution over time with (observed) and without (intrinsic) accounting for the Doppler modulation induced by the relative motion of source and detector, calculated via Equation~\ref{eq:fcw_of_t} with the parameters from \cite{Ming:2025ehy} as reproduced in Table~\ref{tab:G347cand}.
    The inclusion of uncertainties and/or $\dddot f_0$ does not lead to visually appreciable differences.
    Shaded regions highlight the O3 and O4ab data-taking periods.
    \label{fig:outlier_track}
    }
\end{figure}

Figure~\ref{fig:outlier_track} shows the expected frequency evolution in O3 and O4ab data using the parameters reported in this section.

\subsection{Astrophysical interpretation of the candidate}
In this section, we assume the candidate is produced from a rotating NS and infer some of its properties.

First, we use the reported parameters in Table~\ref{tab:G347cand} to estimate a nominal braking index for the source associated with the outlier as a consistency check on its astrophysical interpretation.
For a given spin-down mechanism, the frequency evolution is expected to satisfy $\dot f_0 \propto f^n_0$~\cite{LIGOScientific:2008hqb}, where $n$ is the braking index of the NS.
Using the parameters in Table~\ref{tab:G347cand}, we obtain
\begin{equation}
    \label{eq:brak_idx_outlier}
    n_{\rm outlier} = \frac{(f_0 \pm \Delta^{99\%} f_0) (\ddot f_0 \pm \Delta^{99\%} \ddot f_0)}{(|\dot f_0| \pm \Delta^{99\%} \dot f_0)^2} \sim 18.76 - 19.15\,.
\end{equation}
This is substantially larger than the standard EM- ($n=3$) or CW-dominated ($n=5$) expectations. 
However, this estimate should be interpreted with caution, as the measured frequency derivatives may be affected by the source's motion (e.g., binary~\cite{Leaci:2015bka} or proper-motion~\cite{Covas:2020hcy} effects) or by intrinsic rotational irregularities such as glitches or timing noise~\cite{Parthasarathy:2019txt}.
Using the nominal value in Equation~\eqref{eq:brak_idx_outlier}, the implied third-order frequency derivative would be
\begin{equation}
    \label{eq:fdddot_from_brak_idx}
    \dddot f_0 = (2n-1) \, \frac{\dot f_0 \ddot f_0}{f_0} \approx -3.25 \times 10^{-29}~\mathrm{Hz/s^3}\,.
\end{equation}
We consider this value for reference when discussing prior ranges for our follow-ups, but not as a strict constraint.

Additionally, we can check the rotational and energetic properties implied by the candidate parameters.
First, we assume the standard relation for CW emission from a rotating NS with a non-axisymmetric quadrupole, $f_{\rm rot}$=$f_0/2$ \cite{Riles:2022wwz}, where $f_{\rm rot}$ is the rotational frequency, and similarly for the frequency-time derivatives.
Using the value reported in Table~\ref{tab:G347cand}, we obtain $f_{\rm rot}\approx15.84557$~{Hz}, corresponding to a rotational period of $P_{\rm rot}\approx63.1091$~{\rm ms}, and $\dot{f}_{\rm rot}\approx-1.7992335\times10^{-10}$~{\rm Hz~s$^{-1}$}.
We can compute the rotational-energy loss of the source as $\dot{E}_{\rm rot} = I\Omega\dot{\Omega} = 4\pi^2 I f_{\rm rot}\dot{f}_{\rm rot}$, where $\Omega=2\pi f_{\rm rot}$, and we assume $I=10^{38}~{\rm kg~m^2}$ as the moment of inertia.
Therefore, we estimate the spin-down energy rate via the balance relation $\dot{E}_{\rm rot} = -\dot{E}_{\rm sd}$,
yielding
$\dot{E}_{\rm sd} \approx 1.13\times10^{38}~{\rm erg~s^{-1}}$.

This implies that the candidate from~\cite{Ming:2025ehy} would rotate approximately twice as fast as the fastest previously detected X-ray pulsar in the CCO class~\cite{Gotthelf:2013sa}, with the three fastest such objects having periods of approximately 105, 113, and 424~{\rm ms}.
The corresponding spin-down power inferred here is comparable to that of the Crab pulsar ($\dot{E}_{\rm sd,\,Crab} \approx 4.6\times 10^{38} \rm \, erg~s^{-1}$~\cite{Manchester:2004bp}) and is several orders of magnitude higher than that of the three pulsating CCOs, whose spin-down powers are only of the order of $\dot{E}_{\rm sd,\,CCO}\sim10^{31}-3\times10^{32}~{\rm erg~s^{-1}}$ \cite{Gotthelf:2013sa}.

Two other possible relations between the CW and rotational frequencies give even more extreme inferred NS properties~\cite{Riles:2022wwz}:
If $f_0$=$f_{\rm rot}$ and $\dot{f}_0$=$\dot{f}_{\rm rot}$ (free precession scenario), then $P_{\rm rot} \approx 31.55~{\rm ms}$ and $\dot{E}_{\rm sd} \approx 4.5 \times 10^{38}~{\rm erg~s^{-1}}$.
If $f_0 \approx (4/3)f_{\rm rot}$ and $\dot{f}_0 \approx (4/3)\dot{f}_{\rm rot}$ (dominant quadrupolar r-mode scenario), then $P_{\rm rot} \approx 42.07~{\rm ms}$ and $\dot{E}_{\rm sd} \approx 2.53 \times 10^{38}~{\rm erg~s^{-1}}$.
These alternative interpretations would imply an even faster and more energetic pulsar than in the standard ($f_0 = 2f_{\rm rot}$) scenario.

\section{GW data\label{sec:gw_data}}
We use data from both the O3 run~\cite{KAGRA:2023pio},
which was split into two parts, O3a and O3b,
and from the first two parts of the fourth observing run, called O4a~\cite{LIGOScientific:2025snk} and O4b~\cite{LIGOScientific:2026jgl}, respectively.
The search in~\cite{Ming:2025ehy} used data from O3 and O4a to identify and follow up the outlier, with more details given in Section~\ref{sec:outlier}.

In O3, the LIGO Hanford (H1) and LIGO Livingston (L1) detectors \cite{LIGOScientific:2014pky}, as well as Virgo (V1) \cite{VIRGO:2014yos}, have collected data within two periods, O3a and O3b, separated by a
month-long commissioning break.
O3a occurred from 2019 April 1, 15:00 UTC until 2019 October 1, 15:00 UTC.
O3b ran from 2019 November 1, 15:00 UTC, to 2020
March 27, 17:00 UTC.
More information on detector performance during this period can be found in~\cite{aLIGO:2020wna} for LIGO and~\cite{Virgo:2019juy} for Virgo.
The corresponding calibration uncertainties are reported in~\cite{Sun:2020wke, Sun:2021qcg} for LIGO and~\cite{VIRGO:2021kfv} for Virgo.
We considered inclusion of O3 Virgo data in these follow-ups,
as detailed in section~\ref{sec:virgo}, but concluded that it does not provide additional constraining power.

From O4a and O4b, we also only used H1 and L1 data.
The first part of the run took place between May 24, 2023 15:00 UTC and January 16, 2024 16:00 UTC,
with Virgo not participating.
The three detectors resumed observing for O4b on April 10, 2024, at 15:00 UTC and ended this part of the run on January 28, 2025, at 17:00 UTC.
The LIGO detectors' performance throughout O4a and O4b is described in~\cite{LIGOO4Detector:2023wmz, membersoftheLIGOScientific:2024elc, Capote:2024rmo, LIGO:2024kkz, O4LIGODetector:2026okh}.
The remaining data from the third part of the run (O4c) is expected to be released in December 2026~\cite{dataplan}, and therefore not included in this analysis.
We have not considered V1 data from this run, as the detector joined observations only in O4b, and the difference in sensitivity compared to LIGO has further increased since O3.
See~\cite{VIRGO:2026poi} for more information about the V1 performance in O4.

Furthermore, we have not considered KAGRA data from either O3 or O4, as the detector joined observations only for brief periods at the end of O3 and during O4.

A summary of the start and end times of each run is reported in Table~\ref{tab:start_stop_runtimes}.

\begin{table}[!ht]
    \centering
    \renewcommand*{\arraystretch}{1.4}
    \begin{tabular}{ c  c  c }
    \hline\hline
    Run segment & Start [dd/mm/yyyy (GPS)] & Stop  [dd/mm/yyyy (GPS)] \\
    \hline
    O3a & 01/04/2019 (1238166018) & 01/10/2019 (1253977218)\\
    O3b & 01/11/2019 (1256655618) & 27/03/2020 (1269363618)\\
    O4a & 24/05/2023 (1368975618) & 16/01/2024 (1389456018)\\
    O4b & 10/04/2024 (1396796418) & 28/01/2025 (1422118818) \\
    \hline
    \end{tabular}
    \caption{
    \label{tab:start_stop_runtimes}
    Summary of start and end times of each LVK run segment considered in this work.
    }
\end{table}

\subsection{Data quality investigations\label{sec:data_quality}}

In this section, we discuss data quality checks performed on a narrow frequency band surrounding the candidate, involving run-averaged spectral comparisons and coherence studies in O3 and O4a+b data.

The study by Ming et al.~\cite{Ming:2025ehy} already performed some data-quality checks in O3 and O4a, including the ASDs and known-line lists.
Here, we extend these studies with visualisations of high-resolution spectra, and with several dedicated characterisation tools.

\subsubsection{LIGO data quality\label{sec:ligo_detchar}}

\begin{figure}[!ht]
\centering
\includegraphics[width=\textwidth]{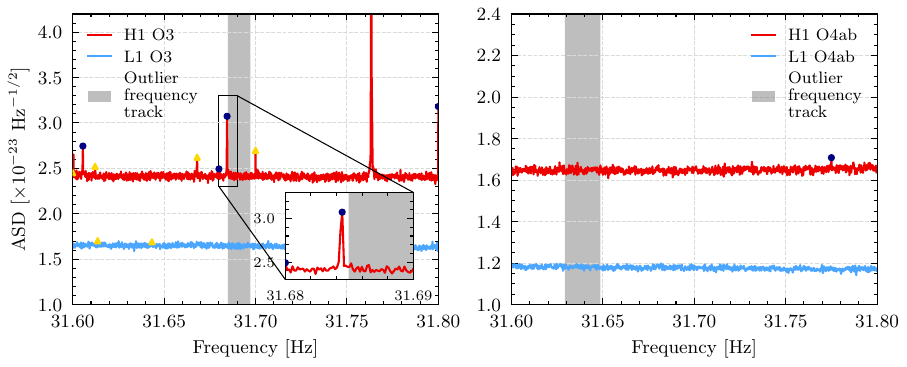}
\caption{
High-resolution amplitude spectral densities (ASDs) of H1 and L1 strain data between $31.6\text{--}31.8$~Hz, averaged over the O3 (left panel) or O4a+b (right panel) runs.
Markers show either vetted (blue circles) or unvetted (yellow triangles) narrow spectral artifacts.
The grey bands show the range of detector-frame frequencies spanned by the G347.3\textminus0.5 candidate in O3 and O4a+b, assuming the nominal parameter values in Table~\ref{tab:G347cand} and Doppler modulations.
\label{fig:runavg_spect}
}
\end{figure}

To assess the quality of the data around the G347.3\textminus0.5 candidate's nominal frequency, we examine high-resolution spectra of H1 and L1 strain data. 
These spectra are calculated from 7200\,s short Fourier transforms (SFTs)~\cite{sftv3} using a noise-weighted averaging procedure detailed in~\cite{O4LIGODetector:2026okh} and implemented in \texttt{LALPulsar}~\cite{lalsuite}.
Figure~\ref{fig:runavg_spect} compares spectra between O3 and O4a+b, focusing on the $31.6\text{--}31.8$~Hz band.
The grey shaded bands in Figure~\ref{fig:runavg_spect} show the range of detector-frame frequencies spanned by the candidate in either O3 or O4a+b, taking into account its presumed spin-down evolution, as given by the parameters in Table~\ref{tab:G347cand}, and Doppler frequency modulations due to the Earth's motion.

For O3 data, Figure~\ref{fig:runavg_spect} reveals a number of narrow spectral artifacts in the vicinity of the candidate, mainly in H1 data.
Some of these are ``vetted'' artifacts associated with known instrumental disturbances, while others are ``unvetted'', meaning no such association has yet been made~\cite{O4LIGODetector:2026okh}.
L1 remains free of artifacts within or near the candidate's frequency band.
The two strongest H1 lines, at $31.68444$~Hz and $31.76333$~Hz (further away from the candidate, and off the scale vertically in Figure~\ref{fig:runavg_spect}) belong to two separate but likely related comb artifacts with exactly $4.0$~Hz spacings.
The origin of these combs is not known.
There are two H1 lines that are close to but do not intersect the outlier frequency track.
These lines are the aforementioned $31.68444$~Hz comb tooth and a weak line at exactly $31.68$~Hz belonging to another comb with $0.16$~Hz spacing, also of unknown origin.
The closest line to the outlier, at $31.68444$~Hz, is narrowly missed by the lower bound of the outlier frequency track during O3 by a few frequency bins, as shown by the insets in Figures~\ref{fig:runavg_spect} and \ref{fig:comparison_spectra_virgo_o3}.

In O4, the band of interest is significantly cleaner,
with the O4a+b averaged spectra not resolving any narrow spectral artifacts near the candidate's frequency band.
The only artifact in the $31.6\text{--}31.8$~Hz band during O4a+b is a relatively weak vetted H1 artifact at $31.775$~Hz, resulting from third-order intermodulation of two temporary calibration lines~\cite{Sun:2020wke, Sun:2021qcg, Wade:2025tgt} at $8.825$~Hz and $11.475$~Hz~\cite{O4LIGODetector:2026okh}, but this is comfortably away from the expected frequency range of the candidate, especially after the putative source is expected to have spun down further since O3.

We further validate the data quality around the G347.3\textminus0.5 candidate by checking for coherence between the primary strain channel and several auxiliary channels, which would be indicative of persistent noise coupling into the strain measurement.
To find auxiliary channels that have high coherence with strain, we parse output generated by \texttt{Fscan} \cite{O4LIGODetector:2026okh} and \texttt{STAMP-PEM} \cite{MeyersThesis} throughout O3 and O4a+b.
We find no evidence of elevated coherence with any auxiliary channels near the frequency of the G347.3\textminus0.5 outlier.

\begin{figure}
\centering
\includegraphics[width=\textwidth]{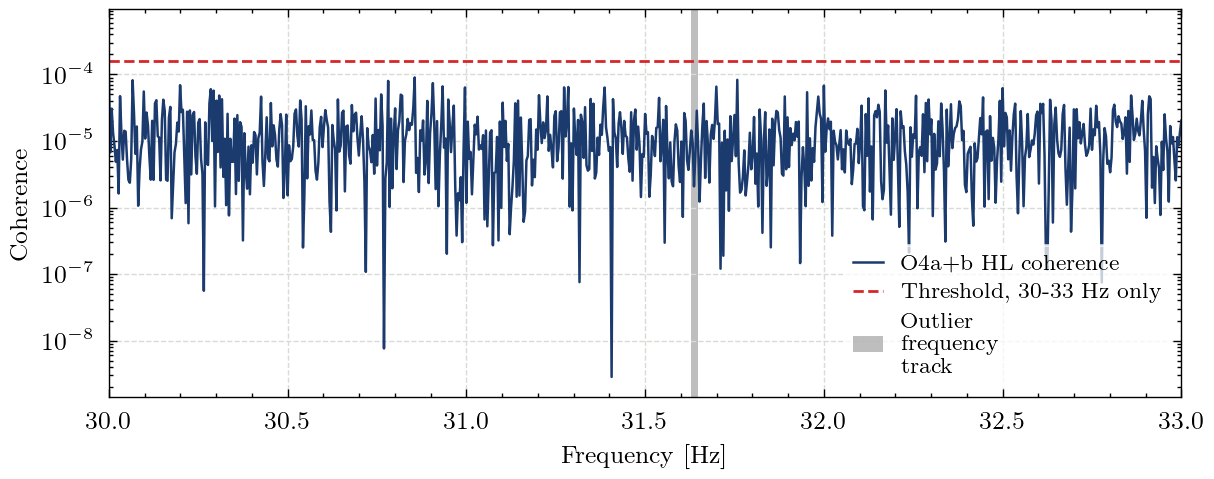}
\caption{
Measured coherence between H1 and L1 from $30\text{--}33$~Hz, using O4a+b data.
The frequency bin resolution is $1/256$~Hz.
The dashed horizontal line is a coherence threshold for identifying outlier frequency bins.
The grey bands show the range of detector-frame frequencies spanned by the G347.3\textminus0.5 candidate during O4a+b, assuming the nominal parameter values in Table~\ref{tab:G347cand} and Doppler modulations.
\label{fig:HL_coherence}
}
\end{figure}

We then use \texttt{Stochmon} \cite{Stochmon} to search for any evidence of correlated noise between the two LIGO detectors.
Pure Gaussian noise is expected to be uncorrelated between detectors;
any narrow peaks or elevated broadband structure in \texttt{Stochmon} spectra would suggest that a correlation is present, providing an entry point for deeper data quality studies~\cite{O4LIGODetector:2026okh}.
Instead of auxiliary channels, \texttt{Stochmon} calculates the coherence between the strain time series from pairs of detectors, modulo a non-physical time shifting that is applied for blinding purposes.
We run \texttt{Stochmon} independently on O3 and O4a+b LIGO data, showing results for O4a+b in Figure~\ref{fig:HL_coherence} as an example.
To decide whether a given level of coherence is statistically significant, a threshold is calculated such that, under the noise hypothesis of uncorrelated Gaussian noise, we expect to find, on average, one frequency bin in the considered band with coherence above this threshold (see Section~2.4 of \cite{O4LIGODetector:2026okh}).
This threshold is shown as a dashed red line in Figure~\ref{fig:HL_coherence}.
Again, we find no evidence for elevated coherence in the vicinity of the G347.3\textminus0.5 candidate in either O3 and O4a+b, and the data appear consistent with the uncorrelated Gaussian noise assumption.

\subsubsection{Considering the inclusion of Virgo data\label{sec:virgo}}

We also investigated whether the inclusion of Virgo (V1) data could improve the constraining power of our analyses.
To do so, we make use of calibrated and cleaned detector data preprocessed through the band-sampled data (BSD)~\cite{Piccinni:2018akm} framework.

\begin{figure}[!ht]
    \centering
    \includegraphics[width=0.75\textwidth]{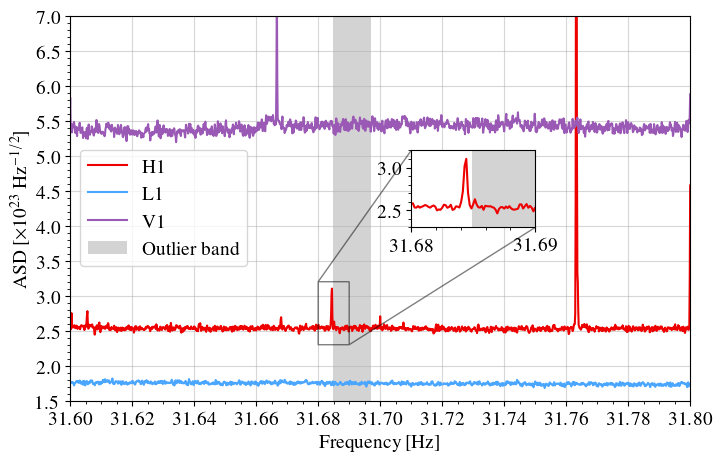}
    \caption{O3 average ASDs, computed every 7200\,s, of the three detectors around the frequency of the outlier. Differences from Figure~\ref{fig:runavg_spect} are related to the additional cleaning performed by the BSD framework~\cite{Piccinni:2018akm} and different averaging procedures.
    The grey-shaded band represents the frequency range spanned by the candidate's track during O3. The inset shows a zoom around the H1 artifact mentioned earlier in Section~\ref{sec:ligo_detchar}, which does not overlap with the outlier's track. 
    The V1 ASD is significantly higher than those of the LIGO detectors, making its contribution to the overall search sensitivity not significant; see the main text for more details.
    A similar scenario occurs for O4b V1 data. \label{fig:comparison_spectra_virgo_o3}}
\end{figure}

Figure~\ref{fig:comparison_spectra_virgo_o3} compares the ASDs of H1, L1, and V1 in the frequency band of interest for the outlier during O3, 
with the V1 ASD generally worse by a factor of $\sim 3.1$ ($\sim 2.1$) times that of L1 (H1).

At the SNR sky location, the antenna pattern functions in O3 averaged over time and polarisations are comparable among detectors: about 0.294 for H1, 0.277 for L1, and 0.288 for V1.
Considering that the O3 duty cycles are also comparable~\cite{KAGRA:2023pio}, sensitivity comparisons among detectors are dominated by ASD variations~\cite{DOnofrio:2024nhu}, implying that the inclusion of V1 data in this analysis, focused on a comparatively faint outlier, would not be more constraining than H1 and L1 data alone. A similar situation holds for O4b.

\section{Confirmation in O3 data\label{sec:o3_search}}

Since we cannot conclusively link the outlier with an instrumental disturbance, we performed three independent analyses testing the CW scenario with the signal model from Equation~\ref{eq:fcw_of_t} using O3 and O4ab data.
Results are reported for the \cwinpy~\cite{cwinpy}, \weave~\cite{Wette2018}, and \pyfstat~\cite{Keitel:2021xeq} pipelines, and here we first cover results on O3 data in Sections~\ref{sec:o3cwinpy}--\ref{sec:o3weave}--\ref{sec:o3pyfstat}, respectively.
All analyses reproduce the candidate consistently.
O4 analyses will be discussed in the following Section~\ref{sec:o4_search}.

Both \cwinpy{} and \pyfstat{} are based on stochastic sampling, while \weave{} is a grid-based search.
All three are based on matched filtering against the Equation~\ref{eq:fcw_of_t} signal model, with \cwinpy{} doing a time-domain analysis and sampling over both frequency-evolution and amplitude parameters, while \weave{} and \pyfstat{} evaluate the frequency-domain $\Fstat$-statistic~\cite{Jaranowski:1998qm,Cutler:2005hc,T0900149} that analytically maximises over amplitude parameters, so they only search over frequency-evolution parameters.
The parameter-space choices for each analysis are discussed in detail below.
All are based on the parameter uncertainties from~\cite{Ming:2025ehy} as reproduced in Table~\ref{tab:G347cand}, but also on computing power and sampler convergence considerations.
Table~\ref{tab:O3pipelines} gives an overview of the configurations covered by each method.

\begin{table}[!ht]
    \centering
    \renewcommand*{\arraystretch}{1.4}
    \begin{tabular}{l c c c}
    \hline\hline
    Search type & \cwinpy{} & \weave{} & \pyfstat{} \\
    \hline
    standard: $\{f_0,\dot f_0,\ddot f_0\}$ & \cmark & \cmark & \cmark\\
    standard + $\dddot f_0$ & \xmark & \xmark & \cmark\\
    standard + varied sky location & \xmark & \cmark & \xmark\\
    standard + varied sky location + $\dddot f_0$ & \xmark & \xmark & \cmark\\
    \hline
    \end{tabular}
    \caption{
    Configurations of O3 parameter space dimensionality considered with each pipeline. See the dedicated subsections for a detailed discussion of the corresponding prior ranges or template banks.
    \label{tab:O3pipelines}
    }
\end{table}

\subsection{\cwinpy{}\label{sec:o3cwinpy}}

The Continuous (gravitational) Wave Inference in Python (\cwinpy) package~\cite{cwinpy} is used to perform a time-domain Bayesian analysis, following the method described by~\cite{Dupuis:2005xv,LIGOScientific:2019xqs}, which we briefly summarise here.

A complex heterodyne is first applied to remove the expected phase evolution of the signal. This accounts for the relative motion of the detector with respect to the source as well as relativistic effects~\cite{Dupuis:2005xv}. The heterodyned data are then passed through a low-pass anti-aliasing filter to suppress the upper sideband generated by the heterodyne and reduce contamination from spectral lines. The filtered data are subsequently down-sampled, producing a time series centred on the expected signal frequency, which has been shifted to 0\,Hz by the heterodyne. If the phase evolution used in the heterodyne perfectly matches the signal, the only signal modulations left in the data are those from the detector antenna patterns (see, e.g., Equation~13 of \cite{Dupuis:2005xv}). If the phase evolution is not a perfect match, the signal in the data will also contain a relatively slowly varying phase modulation (see Equation~8 of \cite{Pitkin:2017qfy}).

Several signal parameters remain unknown, with the amplitude being the primary parameter of interest. Bayesian inference is used to estimate these parameters and to compute the evidence for the signal model.
Posterior sampling is performed using the nested sampling algorithm~\cite{Skilling:2004pqw, Skilling:2006gxv}, as implemented in \texttt{dynesty}~\cite{Speagle:2019ivv} and \texttt{bilby}~\cite{bilby_paper}, with 1024 live points.

For the O3 analysis, we set frequency and spin-down priors to cover the parameter uncertainties reported in Table~\ref{tab:G347cand} enlarged by a factor of 5, fixing the sky location of the source to that of the SNR.
For the other parameters, we use the priors from Appendix~A.2 of~\cite{LIGOScientific:2017hal}, except for $h_0$, for which we use a flat prior within $[0,10^{-20}]$, conservatively reaching much higher than the outlier's estimated amplitude (see Section~\ref{sec:outlier}).

\begin{figure}[!ht]
    \centering
    \includegraphics[width=0.45\textwidth]{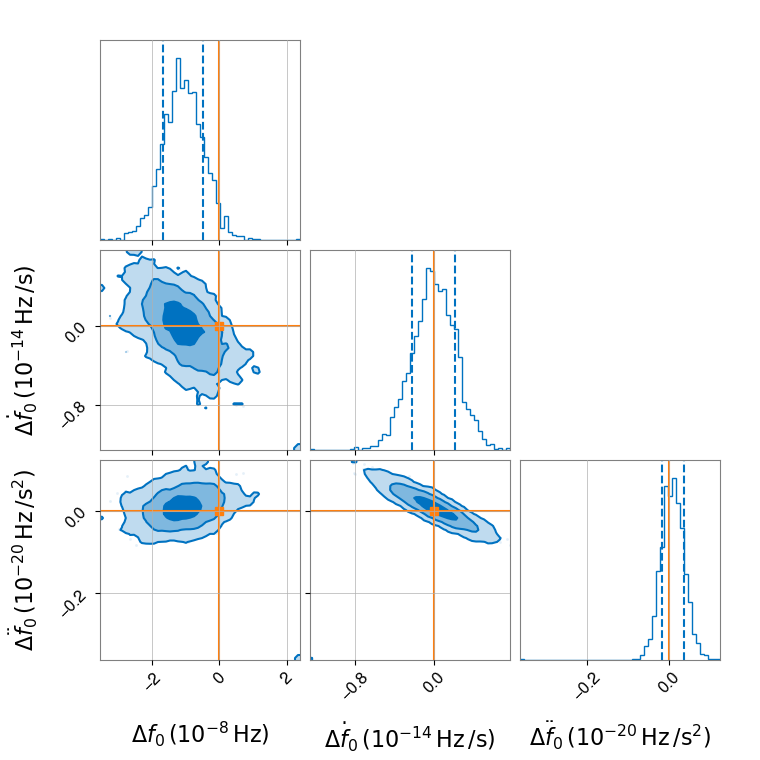}
    \includegraphics[width=0.45\textwidth]{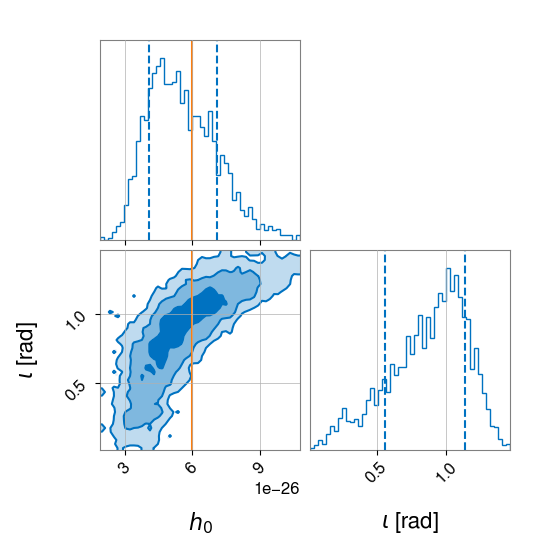}
    \caption{\cwinpy{} posteriors obtained by coherently combining O3 data from H1 and L1. The solid orange vertical lines highlight the outlier parameters from \cite{Ming:2025ehy}, while the vertical dashed blue lines show the 68th percentile of the posteriors. In the left panel, we show posteriors for the frequency-evolution parameters, while the right panel presents the corner plot for signal amplitude and source inclination.
    For greater visual clarity, the less interesting posteriors on $\psi$ and $\phi_0$ are not shown.
    \label{fig:cwinpy_o3_post}
    }
\end{figure}

We perform single-detector and joint analyses, using a likelihood that coherently combines the data from both LIGO detectors, finding consistent results among them.
Our posteriors are in agreement with the results of~\cite{Ming:2025ehy} (Table~\ref{tab:G347cand}).
Figure~\ref{fig:cwinpy_o3_post} shows posteriors from the joint analysis, which is the most constraining, compared with the outlier's.
The $h_0$ posterior is in agreement with the reported $6\times 10^{-26}$~\cite{Ming:2025ehy}, but it shows a degeneracy with the source inclination angle, as expected for a standard CW, and hence is also consistent with a wider range of values, with the 90\% interval of our posteriors lying between $\approx3.4$ and $8.2\times 10^{-26}$.

With these results, we have also computed the base-10 logarithmic Bayes factor comparing the signal versus noise hypotheses. The posteriors presented in Figure~\ref{fig:cwinpy_o3_post} use a wide and flat prior on $h_0$ to be conservative and ensure that posterior results will be dominated by the likelihood given by the data. However, the wide prior creates an unphysical prior volume that gives odds that too quickly favour the noise hypothesis for weak signals \cite{Isi:2017equ}.
To calculate the Bayes factor, we therefore reweight the signal evidences as if they had been calculated by sampling from a log-Uniform prior with bounds $[10^{-33}, 10^{-20}]$, where the lower limit corresponds roughly to a minimum ellipticity of $10^{-12}$ for this source. For the hypothesis test of a coherent signal versus Gaussian noise, we obtain a $\log{}_{10}$ Bayes factor of 3.9, i.e., a coherent signal is $\sim8\,000$ times more likely than noise (considering only the prior volume analysed, not any global trials factors or prior odds).
For the hypothesis test of a coherent signal between both detectors versus incoherent signals in each detector {\it or} Gaussian noise, we obtain a $\log{}_{10}$ Bayes factor of 3.2, i.e., still favouring a signal by a factor of $\sim 1600$, which would be considered as strong evidence (locally). To highlight the effect of prior choices on these values, the equivalent $\log{}_{10}$ Bayes factors using the flat prior were both $\sim 0.11$, which barely support the signal hypothesis.

\subsection{\weave{}\label{sec:o3weave}}

We also analyse the outlier using the \weave{} pipeline~\cite{Wette2018}, a part of \texttt{LALSuite}~\cite{lalsuite}. \weave{} performs coherent or semi-coherent $\Fstat$-statistic searches on a template bank that is based on an optimal parameter-space metric and lattice construction, providing efficient coverage of the parameter space while controlling the maximum loss in signal-to-noise ratio through a user-defined mismatch parameter~\cite{Wette2018, Wette2013, WetteMetrics, WetteLattice}.
We run it here in fully coherent mode.

We used a maximum mismatch of 5\% to perform a fully coherent search over the O3 observing span, with the reference time from Table~\ref{tab:G347cand}. 
The search was centred on the outlier parameters reported in Table~\ref{tab:G347cand}, with search ranges of $(\Delta f_0,\Delta\dot{f}_0,\Delta\ddot{f}_0) = \pm (2 \times 10^{-7}$~Hz, $5 \times 10^{-14}$~Hz\,s$^{-1}$, $7 \times 10^{-21}$~Hz\,s$^{-2}$).
To validate the sky localisation, we also cover a range in $(\alpha,\delta)$ with intervals $\pm(\Delta \alpha,\Delta\delta) = \pm(10^{-4},2\times10^{-4})$ rad.
In addition to the resulting sky grid, we explicitly include the location of G347.3\textminus0.5 as an extra template to ensure that the nominal source location is searched exactly.
For each sky template, the search box in $(f_0,\dot f_0,\ddot f_0)$ is re-centred to account for the change in the detector-frame Doppler modulation associated with the shifted sky position. 
The corresponding offsets are computed using the \texttt{LALSuite} barycentric timing routines and applied to the search-box centre, ensuring that each sky template remains centred on the same underlying source parameters while only the sky position is varied.

In addition, we inject a simulated signal with the same parameters as the reported candidate, except for a different signal frequency $f_0=31$ Hz, and conduct the same search, allowing a direct comparison without contamination from the candidate itself.
The injection signal has amplitude $h_0=6.8\times 10^{-26}$ and $\iota = 1$, consistent with the \cwinpy{} posteriors in Figure~\ref{fig:cwinpy_o3_post}.
For this setup, the same search performed without any injected signal yields a maximum $2\Fstat\sim 24$.
This is below the expected loudest value of $2\Fstat\sim28$ from Gaussian noise for the searched template bank, indicating that the injection band does not contain a loud non-Gaussian artifact.

\begin{figure}[!ht]
    \centering
    \includegraphics[width=\textwidth]{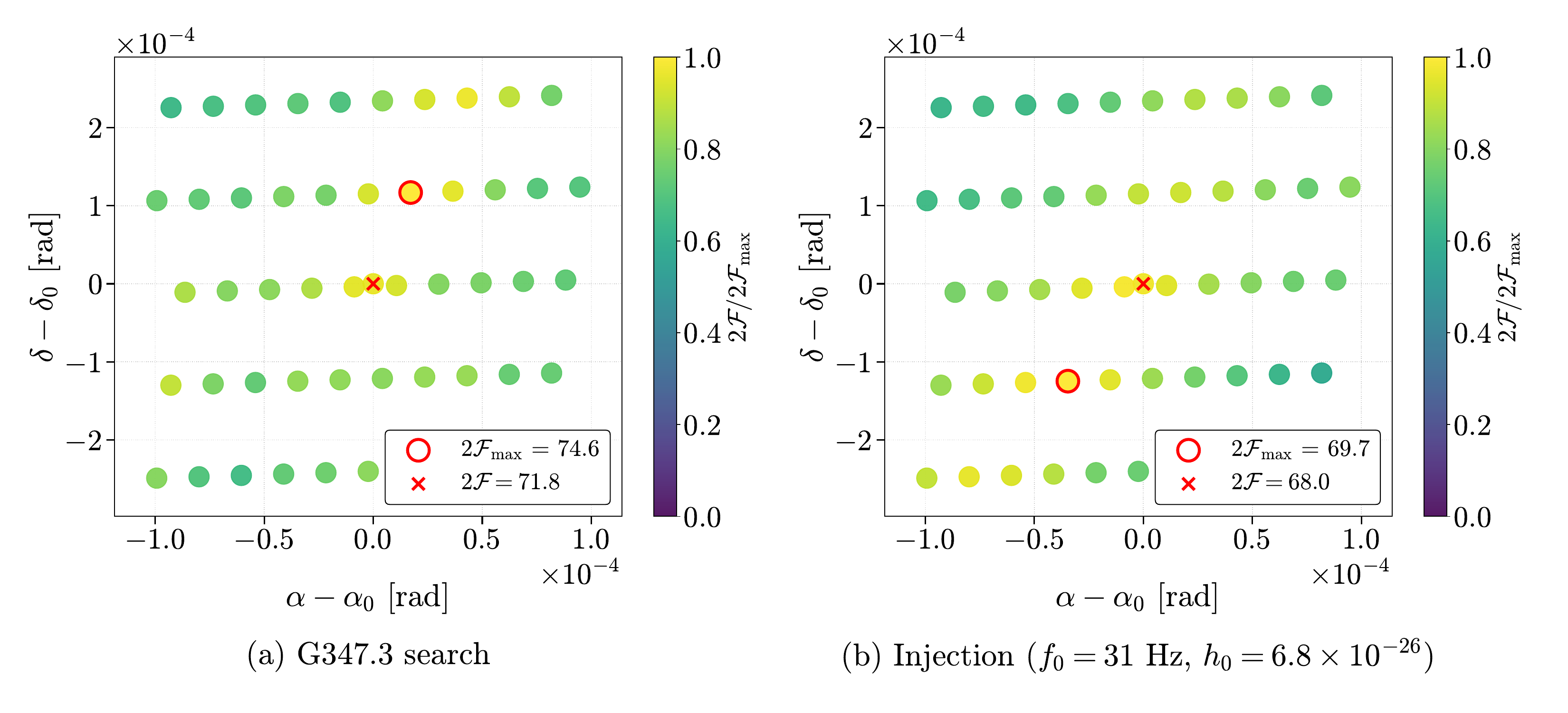}
    \caption{\weave{} normalised $\Fstat$-statistic ratio $2\Fstat/ 2\Fstat_{\mathrm{max}}$ across a sky grid around G347.3\textminus0.5, obtained by coherently combining O3 data from H1 and L1.
    The red cross denotes the nominal G347.3\textminus0.5 sky location, while the red circle marks the sky location corresponding to the maximum detection statistic across the grid, $2\Fstat_{\mathrm{max}}$. 
    The left panel shows the search around the reported G347.3\textminus0.5 outlier parameters, while the right panel shows the corresponding search with an injected signal ($h_0=6.8\times10^{-26}$, $f_0=31$ Hz).
    The maximum $\Fstat$-statistic for the G347.3\textminus0.5 search is $2\Fstat_{\mathrm{max}}=74.6$, located approximately one sky-grid bin from the nominal G347.3\textminus0.5 position, comparable to the offset recovered in the injection test, for which $2\Fstat_{\mathrm{max}}=69.7$.
    \label{fig:weave_o3_sky}
    }
\end{figure}

Figure~\ref{fig:weave_o3_sky} compares the resulting sky maps obtained for the search around the reported outlier parameters and for the injected signal. 
The $\Fstat$-statistic is filtered using the requirement
$
2\Fstat_{\rm L1}>2\Fstat_{\rm H1},
$
since the L1 ASD is lower than the H1 ASD (Figure~\ref{fig:runavg_spect}).
The maximum $\Fstat$-statistic from the G347.3\textminus0.5 search is $2\Fstat_{\max}=74.6$, located approximately one sky-grid template away from the nominal G347.3\textminus0.5 position in both right ascension and declination.
This offset is comparable to that recovered in the injection test, for which $2\Fstat_{\max}=69.7$.

We estimate a local $\pfa\sim 10^{-8}$ by evaluating the probability of obtaining a maximum $2\Fstat$ at least as large as the maximum observed value under the Gaussian-noise hypothesis,
i.e., $\pfa = 1-p(2\Fstat\leq 2\Fstat_{\max})^{N_\mathrm{temp}}$ where \mbox{$p(2\Fstat\leq 2\Fstat_{\max})$} is the $\chi^2$ cumulative distribution function.
The similarity between the sky offsets between the real case and the injection indicates that this feature does not reduce confidence in the candidate based on O3 data.

\subsection{\pyfstat{}\label{sec:o3pyfstat}}

Lastly, we also followed up the G347.3\textminus0.5 candidate using the
\pyfstat{} package~\cite{Keitel:2021xeq, pyfstat}\footnote{The version used for this paper was a development version functionally equivalent to the upcoming release 2.4.0.}, which, like \weave{}, uses the $\Fstat$-statistic, but also allows exploration of parameter space regions with stochastic sampling \cite{Ashton:2018ure, Tenorio:2021njf},  like \cwinpy{}.
Unless stated otherwise, the analyses presented in this section use the \texttt{ptemcee}
sampler \cite{Vousden:2016eeu} 
with a relatively cheap setup of 100 walkers, 3 temperatures, 200 burn-in steps, and 300 production steps, following the recommendations for multi-stage follow-ups in~\cite{Mirasola:2024lcq}.
However, here we perform single-stage fully-coherent analyses.
That these settings are sufficient for robust convergence has been verified with a set of 100 injections; see more information below.
Selected analyses were also repeated using more expensive sampler settings and also with the \texttt{dynesty} nested sampler \cite{Speagle:2019ivv}, 
more recently integrated with \pyfstat{} (\cite{FerrerMartinez:bilbypyfstat_inprep}, in preparation), 
to verify that the results were not dependent on the choice of stochastic sampler.
Both samplers generally yield consistent parameter estimates and maximum detection statistics.

We first searched over the frequency evolution parameters $(f_0,\dot{f}_0,\ddot{f}_0)$, fixing the sky position to that of G347.3\textminus0.5. 
We used uniform priors corresponding to five times the uncertainties reported by \cite{Ming:2025ehy} (Table~\ref{tab:G347cand}), centred on the candidate parameters listed in the same table. 
We performed both single-detector and joint analyses, 
recovering parameter estimates that are consistent across the individual detectors, with the values reported by \cite{Ming:2025ehy}, and with the results obtained by other pipelines.
The maximum detection statistic
is $2\Fstat\approx78$, close to the 81 reported by \cite{Ming:2025ehy} and the 75 from \weave{}.

\begin{figure}[!ht]
    \centering
    \includegraphics[width=\textwidth]{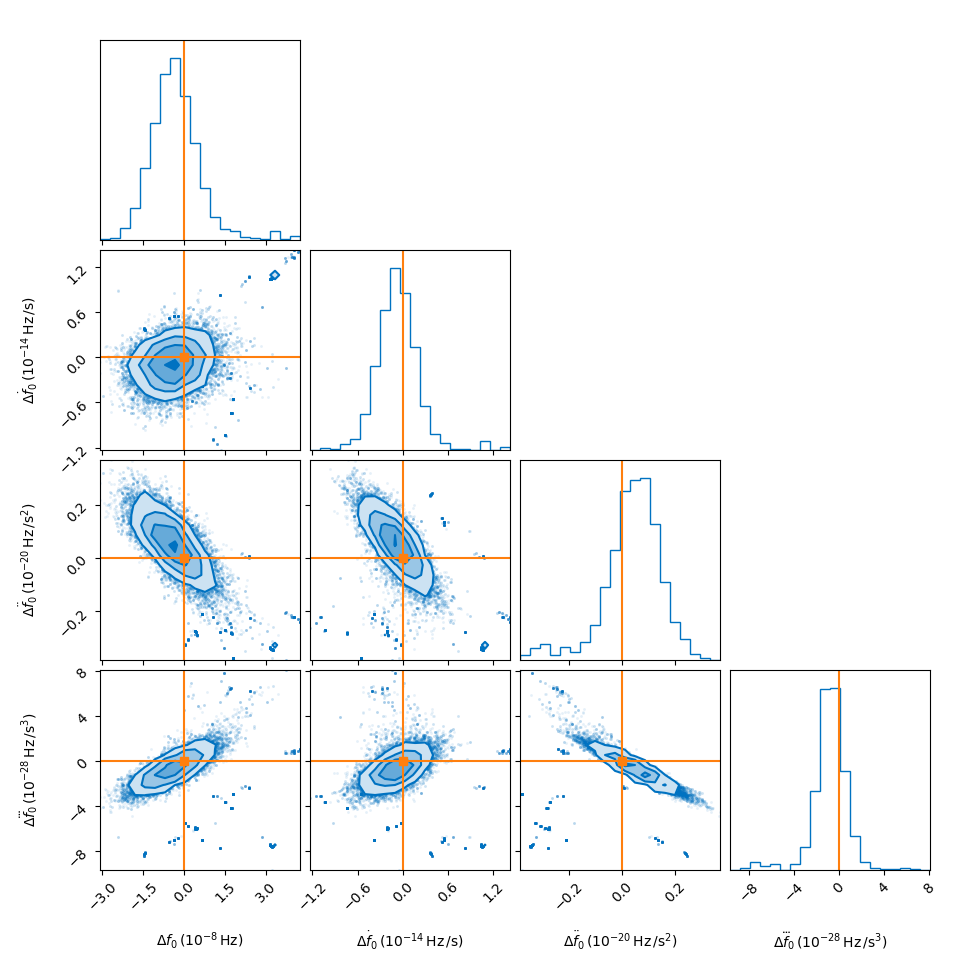}
    \caption{\pyfstat{} results obtained for the O3 joint-detector analysis including $\dddot{f}_0$ while keeping the sky position fixed to that of G347.3\textminus0.5. The panels show the posterior distributions of the parameter offsets, $\Delta f_0$, $\Delta\dot{f}_0$, $\Delta\ddot{f}_0$, and $\Delta\dddot{f}_0$, relative to the candidate values listed in Table~\ref{tab:G347cand}, with zero offsets indicated by the orange solid lines.
    \label{fig:pyfstat_o3_f3dot}
    }
\end{figure}

Similarly to \cite{Ming:2025ehy}, we also investigate the impact of extending the phase model to include a third-order frequency derivative.
To this end, we repeated the analysis including the third-order frequency derivative $\dddot{f}_0$, adopting a fairly wide uniform prior over $\pm10^{-27}\,\mathrm{Hz/s^3}$. 
The posterior distribution, shown in Figure~\ref{fig:pyfstat_o3_f3dot}, remains consistent with $\dddot{f}_0=0$, indicating that the O3 data do not prefer a substantial non-zero third-order frequency derivative.
For the two-detector analysis with this configuration, we found $2\mathcal{F}\approx75$.

\begin{figure}[!ht]
    \centering
    \includegraphics[width=0.5\textwidth]{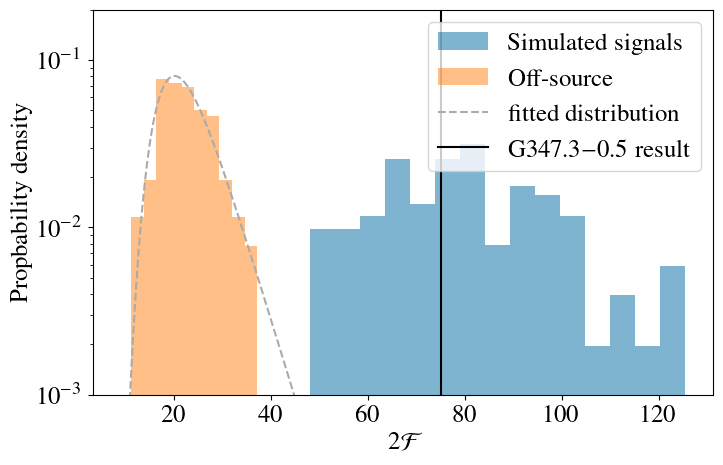}
    \includegraphics[width=0.45\textwidth]{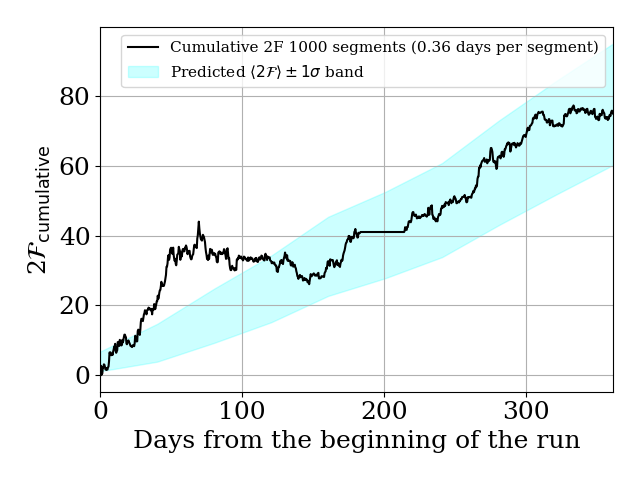}
    \caption{Signal-consistency tests for the \pyfstat{} O3 analysis including $\dddot{f}_0$ with a fixed sky position. Left: distribution of the maximum $2\mathcal{F}$ statistics recovered from 100 off-sourced MCMC runs on the original data (used as a proxy for noise realisations, with a Gumbel distribution fit, see text for details) and 100 simulated signal injections (also off-sourced, see text for details), together with the value recovered for the G347.3\textminus0.5 candidate (vertical line). Right: evolution of the cumulative $2\mathcal{F}$ statistic as a function of the observing time for the loudest template from the main candidate analysis, compared with the prediction for a persistent CW signal. In both cases, the behaviour of the candidate is consistent with the signal hypothesis.
    \label{fig:off_sourcing-2F_evolution}
    }
\end{figure}

To further assess the signal hypothesis, we performed two additional sets of runs as
consistency tests, again including $\dddot{f}_0$ with a fixed sky position for each run. 
Figure~\ref{fig:off_sourcing-2F_evolution} compares the expected 2$\Fstat$ distributions in the absence of a signal and for simulated signals of similar parameters. 
These were estimated from 100 runs with off-sourced sky positions (following \cite{Mirasola:2024lcq}) on the original O3 SFTs as ``noise'' cases, and 100 runs on simulated signal injections into the same O3 SFTs, respectively.
For each realisation, the declination was fixed to that of G347.3\textminus0.5, while
the right ascension was drawn randomly from a uniform distribution subject to an angular separation of at
least $45^\circ$ from the SNR, similarly to the procedure
described in Appendix~B3 of~\cite{LIGOScientific:2026plm}. 
This ensures that both noise and injection studies are performed on data uncorrelated with the candidate. 
In both cases, the same priors on the frequency-evolution parameters were used as in the candidate follow-up.
For the injections, we generate signals using the candidate's frequency-evolution parameters listed in Table~\ref{tab:G347cand}, with the modified right ascension
and with $\dddot f_0$ drawn from a flat distribution in $[-4\times 10^{-29},0]~\mathrm{Hz/s^3}$.
The injection's strength is set by fixing $h_0\approx 8.5\times10^{-26}$ and $\cos\iota\approx0.41$, corresponding to the maximum-likelihood estimators~\cite{Jaranowski:1998qm,T0900149} from our reference O3 \pyfstat{} run (with $\dddot f_0$ included and fixed sky location).\footnote{These amplitude parameter values also lie within the 2D 90\% credible interval of the \cwinpy{} posterior shown in Figure~\ref{fig:cwinpy_o3_post}, and produce a signal of the same signal-to-noise ratio as the $h_0=6\times10^{-26}$ from \cite{Ming:2025ehy} at appropriate $\cos\iota$.}

As shown in Figure~\ref{fig:off_sourcing-2F_evolution}, the $2\Fstat_{\mathrm{max}}$ from the actual outlier analysis lies well outside the expected noise distribution, but is fully consistent with the distribution obtained from the simulated injections.
We estimate a local $\pfa$ by fitting the ``noise'' distribution with a Gumbel function following the discussion in~\cite{Mirasola:2024lcq, Tenorio:2021njf}.
We obtain $\pfa\sim6\times10^{-6}$,
but this does not take into account any trials factor related to the multiple previous search stages performed by~\cite{Ming:2025ehy}, nor to the multiple pipelines used in this work.

We also examined the evolution of the cumulative
$2\Fstat$, at the maximum-likelihood point from our reference run, as a function of the coherent observing time. As shown in Figure~\ref{fig:off_sourcing-2F_evolution}, the measured $2\Fstat$ remains mostly within the predicted $1\sigma$ interval for a persistent CW signal within the observation, except for a small excursion early in the run.

Finally, similarly to the \weave{} analysis, we repeated the analysis allowing
the sky position to vary in addition to the frequency evolution parameters up
to the third-order spin-down, adopting uniform priors covering
$\Delta\alpha = \Delta\delta = \pm 1.01\times10^{-4}$ rad 
around the SNR. 
These would correspond to a width of $\sim\pm10$ bins of $10^4\times (f_0\Tcoh)^{-1}$~\cite{Mirasola:2024lcq} in each sky coordinate.
The recovered sky posterior is consistent with G347.3\textminus0.5.,
while 2$\Fstat_{\max}$ shows a sky offset consistent with the one identified by $\weave$ in Section~\ref{sec:o3weave}.

\subsection{Summary from O3 data}
Three separate follow-up analyses of the O3 data using \cwinpy{}, \weave{}, and \pyfstat{} consistently confirm the G347.3\textminus0.5 candidate from \cite{Ming:2025ehy} to be consistent with the standard CW signal model. Despite employing complementary methodologies (time-domain Bayesian inference, $\Fstat$-statistic grid and stochastic analyses), all pipelines recover signal parameters consistent with those reported by~\cite{Ming:2025ehy}.
The \cwinpy{} analysis finds Bayesian evidence favouring a coherent astrophysical signal over Gaussian noise (log-10 Bayes factor of 3.9).
\weave{} recovers a maximum $2\Fstat$ statistic comparable with that in~\cite{Ming:2025ehy} and with corresponding estimated local $\pfa\sim 10^{-8}$. Results are also consistent with simulations, including the small observed offset in sky location for the local maximum of $2\Fstat$.
\pyfstat{} also demonstrates that the recovered detection statistic and its time evolution are compatible with a persistent CW signal while finding no evidence for a non-zero third-order spin-down, a similar sky offset result to \weave{},
and a local $\pfa\sim 6 \times10^{-6}$.
Overall, none of the O3 analyses identifies features that would disfavour the astrophysical signal hypothesis,
though we stress that the false-alarm estimates do not take into account any trials factors related to the multiple previous search stages performed by~\cite{Ming:2025ehy}, nor to using multiple analysis pipelines.

\section{Looking for the outlier in O4a+b data\label{sec:o4_search}}
We now detail our results obtained on O4a and O4b data.
Similarly as before, results are reported for the \cwinpy{}, \weave{}, and \pyfstat{} pipelines in Sections~\ref{sec:o4cwinpy}--\ref{sec:o4weave}--\ref{sec:o4pyfstat}, respectively.
Parameter space choices are based on O3 results and computational considerations, described below for each pipeline, and summarised in Table~\ref{tab:O4pipelines}.

Following the discussion in Section~6 of~\cite{LIGOScientific:2025kei}, we expect analyses focusing on O4a-only data to have comparable sensitivity to the full-O3 data, or slightly higher.
This is related to a shorter run but with an improved noise level~\cite{LIGOScientific:2025snk,Capote:2024rmo}.
The combined O4ab data are then expected to provide significantly improved sensitivity,
but also require more careful treatment of parameter space size and analysis settings due to the finer resolution required to track a persistent signal over a longer observing time.

Additionally, as can be seen in Figure~\ref{fig:outlier_track}, the candidate parameters evolve significantly in the time between O3 and O4.
The extrapolation following Equation~\ref{eq:fcw_of_t} is done by standard LALSuite~\cite{lalsuite} routines for all pipelines used here.
As described below, \weave{} first extrapolates the parameters to a new reference time to facilitate its grid setup, while \cwinpy{} and \pyfstat{} use the original O3 reference time, and extrapolation is handled internally during the sampling.

\begin{table}[!ht]
    \centering
    \renewcommand*{\arraystretch}{1.4}
    \begin{tabular}{l c c c}
    \hline\hline
    Search type & \cwinpy{} & \weave{} & \pyfstat{} \\
    \hline
    standard: $\{f_0,\dot f_0,\ddot f_0\}$ & \xmark & \cmark & \cmark\\
    standard + $\dddot f_0$ & \cmark & \xmark & \cmark\\
    standard + varied sky location & \xmark & \cmark & \xmark\\
    standard + varied sky location + $\dddot f_0$ & \xmark & \xmark & \cmark\\
    \hline
    \end{tabular}
    \caption{
    Configurations of O4ab parameter space dimensionality considered with each pipeline. See the dedicated subsections for a detailed discussion of the corresponding prior ranges or template banks.
    \label{tab:O4pipelines}
    }
\end{table}

\subsection{\cwinpy{}\label{sec:o4cwinpy}}
The \cwinpy{} analysis on O4 data considers three different datasets: O4a-only, O4ab, O3+O4a.

We first consider O4a-only data, performing the single-detector and joint analyses with the inclusion of a third-order frequency derivative (motivated by the reports of more informative posteriors under this inclusion by \cite{Ming:2025ehy}).
Priors in $f_0,\,\dot f_0,\,\ddot f_0$ are 10 times larger than the O3 priors in Section~\ref{sec:o3cwinpy}, and the  $\dddot f_0$ range is chosen as $\pm10^{-29}$~Hz/s$^3$.

\begin{figure}[!ht]
    \centering
    \includegraphics[width=0.45\textwidth]{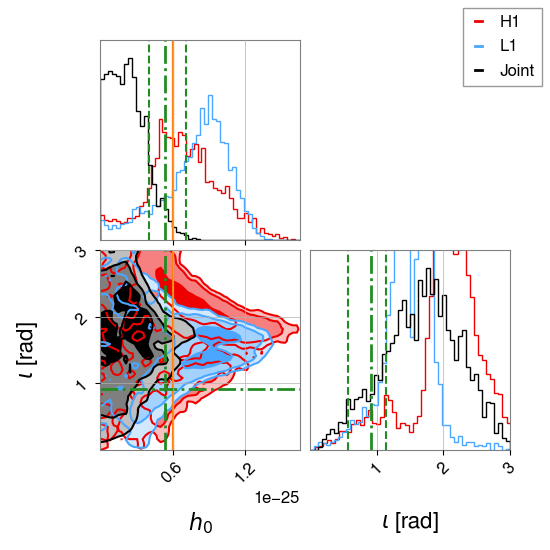}
    \includegraphics[width=0.45\textwidth]{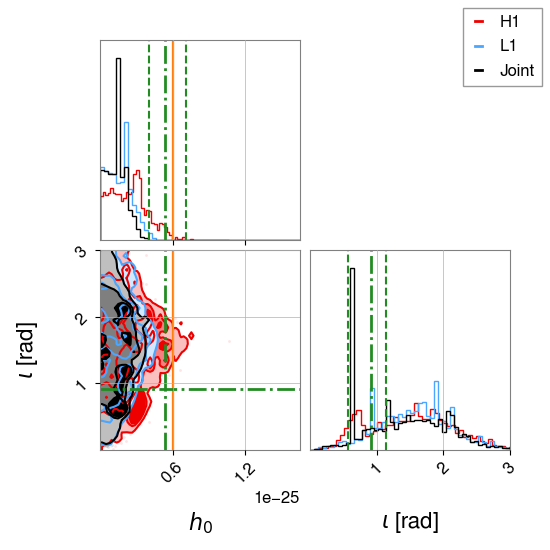}
    \caption{\cwinpy{} posteriors on signal amplitude and the source inclination using O4a (\textit{left}) and O4ab (\textit{right}) data for the single-detector and joint analyses.
    The orange vertical line highlights the outlier's amplitude $h_0$ as reported by \cite{Ming:2025ehy} (see Section~\ref{sec:outlier}),
    while the green dash-dotted lines show both the $h_0$ and $\iota$ posterior medians from the \cwinpy{} O3 analysis in Figure~\ref{fig:cwinpy_o3_post}
    and the green dashed lines (on the 1D posterior histograms only) show the \cwinpy{} O3 68\% intervals.
    \label{fig:cwinpy_o4a_post}
    }
\end{figure}

Unlike O3, the O4a single-detector and joint posteriors differ substantially. While the individual H1 and L1 analyses each show support for a non-zero $h_0$ near the amplitude inferred in O3, the joint analysis (left panel of Figure~\ref{fig:cwinpy_o4a_post}) yields a posterior that is fully consistent with $h_0 = 0$, indicating no evidence for a signal coherent across both detectors. 
Furthermore, the frequency remains unconstrained in all analyses, and the posteriors for its time derivatives exhibit slight multimodal structure that is not consistently reproduced between the single-detector and joint analyses.
As a result, the log-10 Bayes factors (after re-weighting to log-uniform priors as described in Section~\ref{sec:o3cwinpy}) for signal-vs-noise ($-0.23$) and coherent-vs-incoherent ($-0.54$) hypothesis tests are disfavouring the presence of a standard CW signal (following Equation~\ref{eq:fcw_of_t}) in O4a data.

These results are qualitatively consistent with what was reported for the Bayesian follow-up~\cite{Martins:2025jnq} in~\cite{Ming:2025ehy}, where O4a posteriors were consistent with priors and described as ``not convincing'' in terms of O3--O4a consistency.

The coherent combination of O4a and O4b data does not produce more evidence for a signal.
Both the single-detector and joint posteriors are consistent with the priors and significantly disfavour the amplitude estimated in O3 (see the right panel of Figure~\ref{fig:cwinpy_o4a_post}).
The estimated log-10 Bayes factors
($-0.26$ and $-0.54$ for the signal-vs-noise and coherent-vs-incoherent hypothesis tests)
are comparable with those from O4a, showing no support for a signal in the considered dataset.

Lastly, we consider combining O3 and O4a data, but due to the long gap between the data sets we also allow for a less standard scenario in which the source could have glitched \cite{Antonopoulou:2022rpq} before and/or within O4.
We do not include O4b data to contain the computing cost of the analysis.
We use priors as for the O4a-only analysis, performing single-detector and joint analyses again. 
We use uniform priors for the two phase jumps and the within-O4a glitch epoch.
Full posteriors for this search are shown in Appendix~\ref{app:cwinpy}.

Similarly to the O4a-only analysis, we again observe inconsistencies between the different detector choices.
Overall, none show evidence for a phase jump between O3 and O4a.
The combined O3+O4a L1 data support the presence of a signal, with well-constrained frequency parameters and showing a correlation between $\iota$ and $h_0$ as in Figure~\ref{fig:cwinpy_o3_post}.
A within-O4a glitch is disfavoured by the L1 data, as the posterior of the corresponding glitch epoch peaks strongly at the end of the observing run.
On the other hand, the H1-only posteriors show two prominent modes in the frequency parameters, none of them consistent with the L1 results.
H1 results also show a mild bi-modality in the within-O4a glitch-epoch posteriors near the beginning and end of O4a.
The inferred amplitude parameters overlap with those from L1;
the $h_0$ posteriors are peaked around the quoted amplitude in O3, but still have support at $h_0=0$.

The joint H1+L1 inference is driven primarily by the O3 data, with the O4a data contributing comparatively little to the posterior.
As a result, posteriors lock onto a single frequency mode, with $h_0$ peaked again away from zero.
They support a glitch in O4a but without full overlap in the inferred glitch epoch with the single-detector inferences.

Overall, the combination of O3 and O4a data from both detectors increases the log-10 Bayes factors compared to the O4a-only values, and we obtain $2.77$ and $2.43$ for the signal-vs-noise and coherent-vs-incoherent comparisons, respectively. These favour the presence of a signal, but decrease its significance compared to the values from the O3-only data. For a real signal, the addition of more data should increase the evidence rather than decrease it (although the O3 and O4a analysis increases the prior volume with its larger frequency range and glitch priors compared to the O3-only analysis, which partially counteract this), suggesting that these Bayes factors are dominated by the O3 data.

\subsection{\weave{}\label{sec:o4weave}}

We perform three \weave{} searches using the O4 data:

\begin{itemize}
    \item A fully coherent search on O4a with a reference time of GPS \texttt{1379214442}.
    \item A fully coherent search on O4b with a reference time of GPS \texttt{1409520878}.
    \item A two-segment semi-coherent search combining O4a and O4b results, using one coherent segment for each observing run and a reference time of GPS \texttt{1395617439}.
\end{itemize}

The template banks are constructed following the procedure described in Section~\ref{sec:o3weave}, using a maximum mismatch of 5\% for the fully coherent searches and both maximum coherent and semi-coherent mismatches of 5\% for the two-segment search.

The searches are centred on the candidate parameters propagated from the O3 reference time. 
To account for possible deviations from the O3 timing solution, including potential glitches between O3 and O4, we enlarge the search ranges to $\Delta f_0=\pm1\times10^{-4}$\,Hz, $\Delta\dot{f}_0=\pm1.5\times10^{-12}$\,Hz\,s$^{-1}$, and $\Delta\ddot{f}_0=\pm2\times10^{-20}$\,Hz\,s$^{-2}$, while otherwise following the search procedure described in Section~\ref{sec:o3weave}.

The loudest template returned from the O4a search reaches $2\Fstat\approx46.3$, comparable to the value reported by~\cite{Ming:2025ehy}.
The corresponding values for the O4b and two-segment semi-coherent O4ab searches are $2\Fstat\approx45.5$ and $2\hat{\Fstat}\approx33.0$.
These are much lower than the O3 result, and given the large number of templates searched, a small number of templates with $2\Fstat$ values in this range is expected from pure Gaussian noise.
The local $\pfa$ for each of the three O4 searches is approximately unity, indicating that values at least as large as those observed are readily obtained under the Gaussian-noise hypothesis.

\begin{figure}[!ht]
    \centering
    \includegraphics[width=\textwidth]{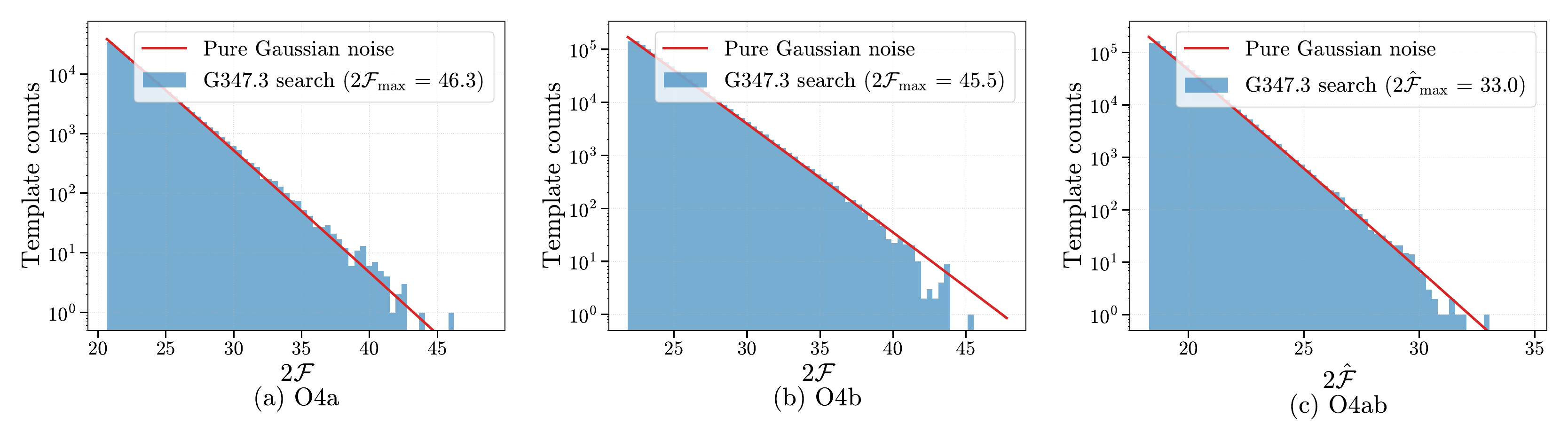}
    \caption{Distributions of per-template $2\Fstat$ results from the \weave{} searches around the G347.3\textminus0.5 outlier. Panels show (a) the fully coherent O4a search, (b) the fully coherent O4b search, and (c) the two-segment semi-coherent O4ab search.
    The histograms show the $2\Fstat$ values of the top 200,000 templates returned by each search. 
    The red lines show the expected template-count distributions for pure Gaussian noise, assuming a $\chi^2$ distribution with four degrees of freedom for the fully coherent searches and the corresponding Gaussian-noise distribution of the averaged semi-coherent statistic $2\hat{\Fstat}$ for the O4ab search.
    \label{fig:weave_o4_template_vs_2F}
    }
\end{figure}

To quantify this, Figure~\ref{fig:weave_o4_template_vs_2F} compares the template $2\Fstat$ distributions obtained from the searches around the G347.3\textminus0.5 outlier with the theoretical template-count distribution expected under the Gaussian-noise hypothesis. 
Panels (a), (b), and (c) correspond to the fully coherent O4a search, the fully coherent O4b search, and the two-segment semi-coherent O4ab search, respectively. 
For the semi-coherent search, we use the averaged semi-coherent detection statistic $2\hat{\Fstat}$, obtained by averaging over the $\Nseg=2$ coherent segments, similarly to the search in~\cite{Ming:2025ehy}.
For all three searches, the observed distribution closely follows the Gaussian noise expectation. No statistically significant excess of high-$2\Fstat$ templates is observed, indicating that data near the outlier parameters appear consistent with Gaussian noise in all three searches.

\begin{figure}[!ht]
    \centering
    \includegraphics[width=\textwidth]{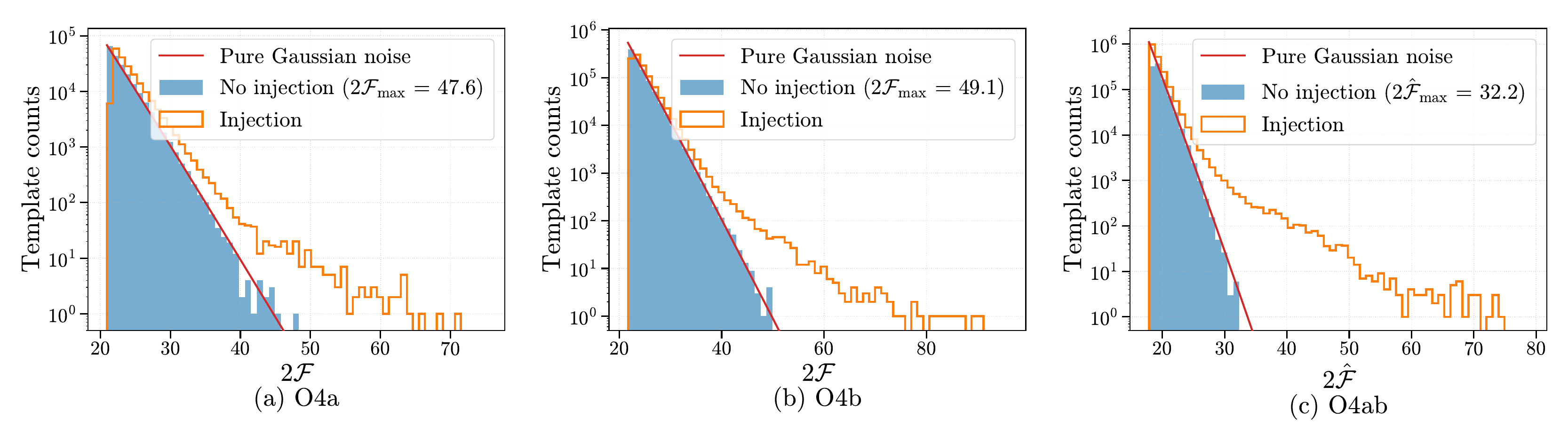}
    \caption{Distributions of per-template $2\Fstat$ results from the \weave{} searches at 31.85~Hz without (blue) and with (orange) an injected signal ($h_0=6.8\times10^{-26}$). Panels show (a) the fully coherent O4a search, (b) the fully coherent O4b search, and (c) the two-segment semi-coherent O4ab search. 
    The red lines show the same theoretical template-count distributions expected under the Gaussian-noise hypothesis as in Figure~\ref{fig:weave_o4_template_vs_2F}.}
    \label{fig:weave_o4a_inj_template_vs_2F}
\end{figure}

To illustrate the expected behaviour in the presence of a CW signal, Figure~\ref{fig:weave_o4a_inj_template_vs_2F} compares searches performed with and without an injected signal at 31.85\,Hz (using the same $h_0$ and $\iota$ as in Section~\ref{sec:o3weave}), again for the three search configurations
(fully coherent O4a search, fully coherent O4b search, and two-segment semi-coherent O4ab search). 
In all three searches, the distributions without an injected signal closely follow the Gaussian expectation, whereas the searches with an injected signal exhibit a pronounced excess of high-$2\Fstat$ templates.
Such an excess is not observed in the G347.3\textminus0.5 searches using O4 data.

The consistency of the G347.3\textminus0.5 searches with the Gaussian-noise expectation in the fully coherent O4a and O4b searches, as well as the two-segment semi-coherent O4ab search, provides no evidence that the candidate corresponds to a truly persistent CW signal in the O4 data following the standard signal model of Equation~\eqref{eq:fcw_of_t}.
The enlarged parameter space searched in O4 accommodates moderate parameter offsets from the O3 candidate parameters, including those that could result from a modest glitch between O3 and O4. 
However, if the source deviated significantly from the assumed signal model, e.g. if it experienced a sufficiently large glitch, or remained in a prolonged post-glitch recovery phase with an exponential relaxation \cite{Antonopoulou:2022rpq} that deviates significantly from the canonical signal model, the recovered $2\Fstat$ would be substantially reduced, and the signal could remain undetected despite being astrophysical in origin.

\subsection{\pyfstat{}\label{sec:o4pyfstat}}
Lastly, we conduct \pyfstat{} analyses on O4a and O4ab data (combined fully coherently) to investigate the same scenarios as in Section~\ref{sec:o3pyfstat}.
The extrapolation from the O3 reference time and the longer observing time of O4ab make the effective parameter space size
(see discussion in Section IV of \cite{Ashton:2018ure})
more challenging for stochastic sampling than the original same-data follow-up in O3, especially when allowing for a third-order spin-down term.
We present O4a results using the same baseline \texttt{ptemcee} sampler configurations as before for O3, while for O4ab, we transition to a more computationally demanding configuration (500 walkers, 3 temperatures, 0 burn-in, and 1000 production steps).
The robustness of these settings is again confirmed with injections
and with spot-checks using more expensive settings and the alternative \texttt{dynesty} sampler.

We first performed an analysis on O4a data with $\dddot{f}_0=0$ and fixing the sky location to that of G347.3\textminus0.5.
We used Gaussian priors with means set to those of the O3 posteriors, enlarging the standard deviations to twice those of the corresponding O3 posteriors.
For O4a, the maximum recovered detection statistic is $2\Fstat \approx 31$, which is lower than those found by~\cite{Ming:2025ehy} and \weave{} on those data, and it mildly increases to $2\Fstat\approx 33$ with the combined O4ab data.

\begin{figure}[!ht]
    \centering
    \includegraphics[width=0.45\textwidth]{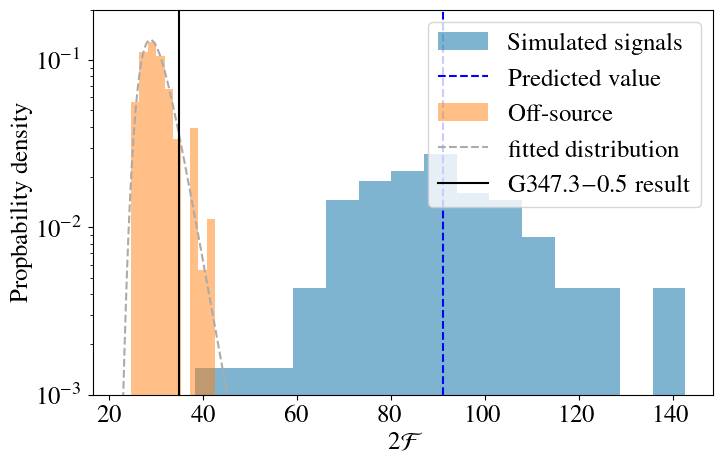}
    \includegraphics[width=0.45\textwidth]{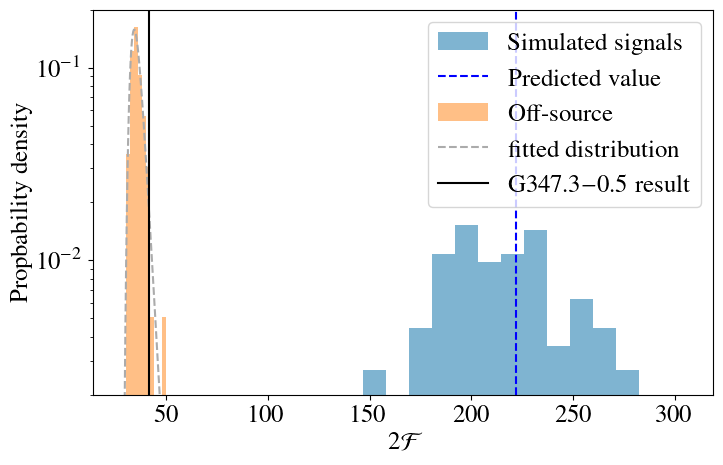}
    \caption{Same as in the left panel of Figure~\ref{fig:off_sourcing-2F_evolution} for the O4a (\textit{left}) and O4ab (\textit{right}) \pyfstat{} analyses.
    Additionally, the predicted $2\Fstat$ from the maximum-likelihood parameters of our O3 reference result (fixed sky, $\dddot f_0$ included)
    is shown for each data set as a dashed vertical line,
    taking into account the relative run durations and ASDs.
    \label{fig:off_sourcing-O4a}
    }
\end{figure}

We then included a non-zero $\dddot{f}_0$ with a Gaussian prior centred at zero with a standard deviation of $4 \times 10^{-29}$~Hz/s$^3$. 
This width conservatively includes the nominal value from the braking index estimate in Equation~\ref{eq:fdddot_from_brak_idx} while not being too wide for robust convergence, as checked with the same set of simulated signals from Section~\ref{sec:o3pyfstat} applied to O4a and O4ab data
(defined at the O3 reference time
and extrapolated to O4 the same way as the candidate itself).
The joint analyses recovered maximum detection statistics
of $2\Fstat \approx 33$ for O4a and $2\Fstat \approx 41$ for O4ab, with posteriors resembling priors. 
In Figure~\ref{fig:off_sourcing-O4a} we show the expected noise distributions (estimated via off-sourcing, as described in Section~\ref{sec:o3pyfstat}) and the distributions of injection recoveries.
Both the O4a and O4ab results are completely in agreement with the noise distribution.
The O4a result is close to, but still below, the low-$2\Fstat$ tail of the signal distribution, which slightly overlaps with the noise one.
The O4ab result, on the other hand, falls significantly short of even the worst-recovered signal injection.
Similarly to the O3 analysis, we estimate a local $\pfa$ by fitting the off-source ``noise'' results with a Gumbel distribution.
We obtain $\pfa\sim11\%$ with O4a data, while the O4ab analysis returns $\pfa\sim5\%$, neither providing evidence for a significant deviation from noise.

Using the same configuration described above, we also performed analyses with uniform priors on $\alpha$ and $\delta$, with bounds set to $\pm 2\sigma$ around the means of their respective O3 posteriors.
The maximum recovered detection statistic is still lower than what was reported by \weave{} in Section~\ref{sec:o4weave}.
By changing to a more computationally demanding configuration of 500 walkers, 5 temperatures, and 1000 steps, we obtain $2\Fstat_{\max} \approx40$ for O4a in this sky-coverage setup, almost matching the values from~\cite{Ming:2025ehy} and \weave{}.
The inclusion of O4b in this last configuration yields a lower $2\Fstat_{\max} \approx32$.
Both values again fall well short of the expected $2\Fstat$ and of simulation results.

Consistent results, across the different prior choices, are obtained with the \texttt{dynesty} sampler.
We have also investigated several additional prior configurations with different widths and/or shapes (uniform/Gaussian), and also performed semi-coherent analyses.
In all cases, results remain consistent:
the recovered maximum $2\Fstat$ is significantly lower than in O3 data and than what is obtained from simulated signal injections with parameters similar to the candidate's.
The highest $2\Fstat_{\max} \approx50$ recovered among all the trials is reported by an O4ab search where F3 is not included, but we set priors on the sky location around the SNR's, slightly higher than what is reported by Weave, but still significantly below expectations.

\subsection{Summary from O4a+b data}
In contrast to the results presented in Section~\ref{sec:o3_search} based on O3 data, O4a and O4b do not show support for a CW signal consistent with the O3 outlier reported by~\cite{Ming:2025ehy}.
This strengthens and extends the original study's findings in O4a, where they already found reduced evidence and unconvincing phase-parameter consistencies with O3, though they did not rule out the candidate.
Although the three pipelines, \cwinpy{}, \weave{}, and \pyfstat{}, explore several scenarios (e.g., inclusion of higher-order spin-down terms, sky-position offsets, glitches before and within O4a), their conclusions are consistent: no significant coherent signal is recovered in O4a and O4b under the standard CW model assumption and the O3-informed priors.

More specifically, the \cwinpy{} log-10 Bayes factor comparing the coherent astrophysical signal and Gaussian-noise hypotheses decreases from 3.9 in O3 to $-0.22$ using O4a alone and $-0.26$ using the combined O4ab dataset. When O3 and O4a are analysed jointly, allowing for a glitch between the observing runs and an additional glitch within O4a, the signal hypothesis is again favoured, with a log-10 Bayes factor of 2.77. However, this value remains lower than that obtained from O3 alone. For a persistent signal following the assumed model (with or without glitches), the addition of more data would be expected to increase, rather than decrease, the Bayes factor, indicating that the remaining support is primarily driven by O3 data.

For the $\Fstat$-statistic based analyses, we can calculate the expected change in the detection statistic using the O3 maximum-likelihood parameters, along with the observing times and ASDs for each run.
Via \pyfstat{}, this yields $2\Fstat\approx91$ for O4a and $\approx222$ for O4ab, compared to the $\approx77$ from O3.

Compared to these expectations,
the \weave{} searches find no evidence for a persistent signal in O4. The loudest candidates reach $2\mathcal{F}\approx46$ in O4a, $2\mathcal{F}\approx46$ in O4b, and $2\hat{\mathcal{F}}\approx33$ in the semi-coherent combined O4ab search.
These values are substantially lower than those obtained in O3 and than the O4 predictions, and are fully consistent with the Gaussian-noise expectations, with $\pfa \sim 1$ for all three searches and no statistically significant excess of high-statistic templates.

The \pyfstat{} analyses lead to the same conclusion. Depending on the prior configuration, the maximum recovered detection statistic ranges from $2\mathcal{F}\approx31$ to $40$ for O4a, while the inclusion of O4b only mildly increases the maxima to $\approx32$ to 50. In all configurations, the recovered values are significantly lower than those obtained in O3 and than expected. The posteriors remain largely prior-dominated, and the recovered statistics are consistent with the noise distribution, providing no evidence for a persistent signal.
For our standard run, i.e., including a non-zero $\dddot f_0$ and with fixed sky position, we obtain a local $\pfa\sim11\%$ with O4a data, while the O4ab analysis returns $\pfa\sim5\%$ (neither accounting for trials factors), showing no evidence for a deviation from noise.

\section{Conclusions\label{sec:conclusions}}
This paper details a multi-pipeline follow-up analysis for the outlier identified in an Einstein@Home directed search~\cite{Ming:2025ehy} for CW signals from the G347.3\textminus0.5 supernova remnant.
Initially identified in LIGO O3 and O4a data~\cite{Ming:2025ehy}, we here report our findings including the newly released O4b dataset.

We report on extended data-quality studies to investigate potential correlations between the outlier and disturbances affecting the detectors,
covering O3 and O4a+b data.
Even with more in-depth studies, our results do not relate the candidate to known artifacts nor to correlated noise between detectors, in agreement with~\cite{Ming:2025ehy}.
As an additional outcome of this study, we concluded that including Virgo data in this dedicated follow-up would not have a higher constraining power than LIGO data alone.
This is related to the ASD ratio at the frequency of interest.

We have also performed dedicated follow-ups in both O3 and O4a+b data with three different pipelines: \cwinpy{}~\cite{cwinpy}, \weave{}~\cite{Wette2018}, and \pyfstat{}~\cite{Keitel:2021xeq}
and present detailed results including Bayesian posteriors.
This ensures that the results are robust to the choice of analysis pipeline and are not driven by any single set of specifications, such as data preprocessing methods or detection statistics.

Taken together, our results do not provide evidence that the candidate is consistent with a standard CW signal persisting across the full available data. 
While the outlier can be robustly recovered in O3 data with multiple analysis methods, the absence of a clear recovery in O4a by~\cite{Ming:2025ehy} and in both O4a and O4b across all three follow-up pipelines used in this work disfavours the interpretation of a long-lived, phase-coherent CW signal under the assumptions adopted in this analysis.

At the same time, these findings do not necessarily exclude more complex signal scenarios that fall outside the standard CW framework.
Future investigations could therefore further explore models that relax these assumptions, beyond the small set of initial tests of pulsar glitch scenarios and semi-coherent O4a+b combinations we have included in this work.

We also emphasise that the G347.3\textminus0.5 candidate reported by~\cite{Ming:2025ehy}
has served as a highly useful test case for follow-up studies of CW candidates more generally.
Lessons learned from this exercise will be of value in working towards the first confident detection of a CW source,
a long-standing science goal that would provide novel insights into the physics and astrophysics of NSs~\cite{Jones:2024npg,Owen:2025ata,Owen:2025ecd}.

\ack{
We thank the AEI Hannover CW team (J.~Ming et al.) and our colleagues in the LIGO--Virgo--KAGRA collaboration for helpful discussions about the candidate.
We thank Andrew L. Miller and the AEI Hannover CW team for helpful feedback on the manuscript.
This material is based upon work supported by NSF's LIGO Laboratory, which is a major facility fully funded by the National Science Foundation.
The authors are grateful for computational resources provided by the LIGO Laboratory and supported by National Science Foundation Grants PHY-0757058 and PHY-0823459.
Part of this work has been done on computing nodes provided and managed by the Rome Virgo group and the computational resources and services provided by Advanced Research Computing at the University of Michigan, Ann Arbor. 

\noindent This document has been assigned LIGO document number \href{https://dcc.ligo.org/P2600403/}{P2600403}.
}

\funding{
L.~Mirasola, M.~Carrio, M.~Ferrer-Martinez, D.~Keitel, O.~J.~Piccinni, I.~Prohens, A.~Calafat, R.~Jaume, I.~La~Rosa, J.R.~Mérou, A.M.~Sintes and R.~Tenorio are supported by the Universitat de les Illes Balears (UIB)  with funds from the Programa de Foment de la Recerca i la Innovació de la UIB 2024--2026 (supported by the yearly plan of the Tourist Stay Tax ITS2023-086); the Spanish Agencia Estatal de Investigación grants PID2022-138626NB-I00, RED2024-153978-E, RED2024-153735-E, funded by MICIU/AEI/10.13039/501100011033 and the ERDF/EU; and the Comunitat Autònoma de les Illes Balears through the Conselleria d'Educació i Universitats with funds from the ERDF (SINCO2022/18146-Plataforma HiTech-IAC3-BIO) and by COST action SCALES CA24139, supported by COST (European Cooperation in Science and Technology).
O.~J.~Piccinni is supported by the Spanish Ministerio de Ciencia, Innovacion y Universidades Ramon y Cajal, RYC2023-044489-I funded by MCIN/AEI/10.13039 /501100011033 and the FSE+ and cofinanced by the Universitat de les Illes Balears (UIB).
M. Bejger and A. Krolak are partially supported by the Polish National Science Center grant no. 2023/49/B/ST9/02777.
D.H.T.~Cheung, A.M.~Knee, and K.~Riles are partially supported by the National Science Foundation Grant PHY-2408883.
B.~Rajbhandari was supported in part by NSF grant PHY-2450793.
S.~Safi-Harb acknowledges support from the Natural Sciences and Engineering Research Council of Canada (NSERC) through the Canada Research Chairs and Discovery Grants programs, as well as from the Canadian Space Agency.
K.~Wette is supported by the Australian Research Council Centre of Excellence for Gravitational Wave Discovery (OzGrav), project number CE230100016.
J. T. Whelan was supported by NSF grant PHY-2409745 and by a Royal Society Wolfson Visiting Fellowship. F.~Amicucci is supported by the BRIDGE project, funded within the SAP\_RICERCA\_2024 research programme.
}

\data{
Analysis configurations and results are available upon reasonable request to the authors.
}

\appendix
\section{Additional \cwinpy{} posteriors\label{app:cwinpy}}
\begin{figure}[!ht]
    \centering
    \includegraphics[width=\textwidth]{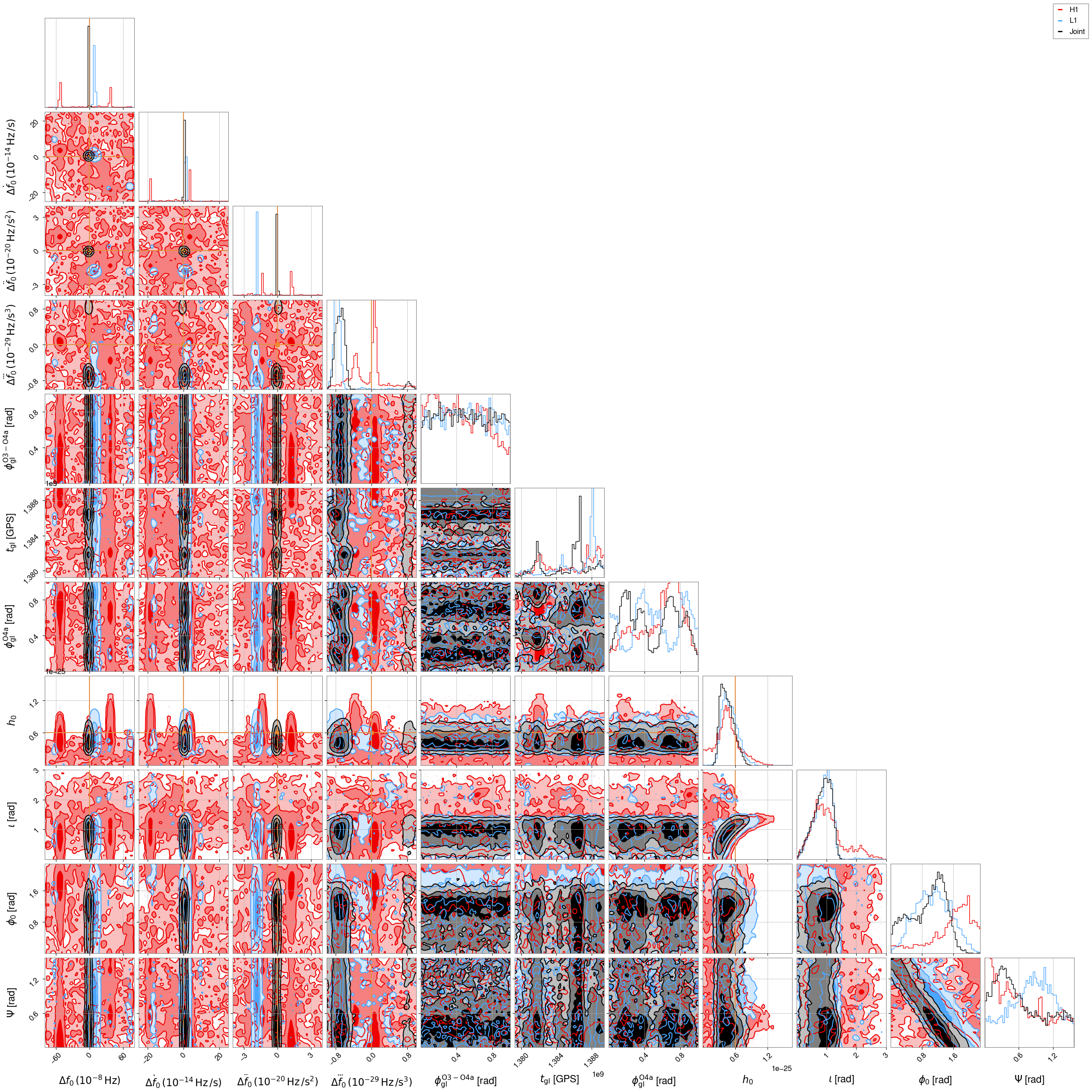}
    \caption{Full \cwinpy{} posteriors using the combined O3+O4a data; see the discussion in Section~\ref{sec:o4cwinpy} for more information. 
    In addition to the standard CW parameters introduced in Section~\ref{subsec:signal}, we here infer the phase of a possible glitch between O3 and O4a ($\phi_{\rm gl}^{\rm O3-O4a}$), the phase of a possible intra-O4a glitch ($\phi_{\rm gl}^{\rm O4a}$), and the latter's epoch ($t_{\rm gl}$).
    The orange vertical lines show the values reported by~\cite{Ming:2025ehy} in Table~\ref{tab:G347cand}.}
    \label{fig:o3o4a_cwinpy}
\end{figure}

We show in Figure~\ref{fig:o3o4a_cwinpy} the full corner plot of the \cwinpy{} analysis on coherently combined O3+O4a data,
allowing for possible NS glitches between O3a and O4 and within o4a.
See Section~\ref{sec:o4cwinpy} for more information on the analysis setup and the interpretation of these results.

\bibliographystyle{iopart-num}
\bibliography{bibliography}

\end{document}